\documentclass{sig-alternate-10pt}
\usepackage{subfigure}
\usepackage{float}
\usepackage{multirow}
\usepackage{graphicx}
\usepackage{multirow}
\usepackage{multicol}
\usepackage{rotating}

\begin{document}

\title{Towards Realistic Vehicular Network Modeling Using Planet-scale Public Webcams}
%
%
%
%
%

\numberofauthors{6} 
%
\author{
%
%
\alignauthor
Gautam S. Thakur$^{\ddag  \S}$ \and Pan Hui$^{\ddag}$  \\ \and Hamed Ketabdar$^{\ddag}$ \and Ahmed Helmy$^{\S}$ \and   \\
                    \affaddr{$^{\S}$CISE, University of Florida, Gainesville,   $^{\ddag}$Deutsche Telekom Laboratories, Berlin}\\\\
             \email{gsthakur@cise.ufl.edu, pan.hui@telekom.de, hamed.ketabdar@telekom.de, helmy@cise.ufl.edu}
}


\maketitle
\begin{abstract}
Realistic modeling of vehicular mobility has been particularly challenging due to a lack of large libraries of measurements in the research community.  In this paper we introduce a novel method for large-scale monitoring, analysis, and  identification of spatio-temporal models for vehicular mobility using the freely available online webcams in cities across the globe. We  collect vehicular mobility traces from 2,700 traffic webcams in 10 different cities for several months and generate a mobility dataset of 7.5 Terabytes consisting of 125 million of images. To the best of our knowledge, this is the largest data set ever used in such study. To process and analyze this data, we propose an efficient and scalable algorithm to estimate traffic density  based on background image subtraction. Initial results show that at least 82\% of individual cameras with less than 5\% deviation from four cities follow Loglogistic distribution and also 94\% cameras from Toronto  follow gamma distribution. The aggregate results from each city also demonstrate that Log-Logistic and gamma distribution pass the KS-test with 95\% confidence.  Furthermore,  many of the camera traces exhibit long range dependence, with self-similarity evident in the aggregates of traffic (per city). We believe our novel data collection method and dataset provide a much needed contribution to the research community for realistic modeling of vehicular networks and mobility.

\end{abstract}




\section{Introduction}

Research in the area of vehicular networks has increased dramatically in recent years. With the proliferation of mobile networking technologies and their integration with the automobile industry, various forms of vehicular networks are being realized. These networks include vehicle-to-vehicle, vehicle-to-roadside, and vehicle-to-roadside-to-vehicle architectures. Realistic modeling, simulation and informed design of such networks face several challenges, mainly due to the lack of large-scale community-wide libraries of vehicular data measurement, and representative models of vehicular mobility.

Earlier studies in this area have clearly established a direct link between vehicular density distribution and the performance \cite{1040698,898398} of vehicular networks primitives and mechanisms, including broadcast and geocast protocols\cite{Bai:2009:SVV:1614269.1614278}. Although good initial efforts have been exerted to capture realistic vehicular density distributions, such efforts were limited by availability of sensed vehicular data\cite{Yeo:2006:CCR:1129582.1129588}. Hence, there is a real need to conduct vehicular density modeling using larger scale and more comprehensive data sets. Furthermore, commonly used assumptions, such as exponential distribution\cite{4346442} of vehicular inter-arrival times\cite{Bai:2009:SVV:1614269.1614278}, have been used to derive many theories and conduct several analyses, the validity of which bears further investigation.

In this study, we provide a novel framework for the systematic monitoring, measurement, analysis and modeling of vehicular density distributions at a large scale. To avoid the limitations of sensed vehicular data, we instead utilize the existing global infrastructure of tens of thousands of video cameras providing a continuous stream of street images from dozens of cities around the world. Millions of images captured from publicly available traffic web cameras are processed using a novel density estimation algorithm, to help investigate and understand the traffic patterns of cities and major highways. Our algorithm employs simple, scalable, and effective background subtraction techniques to process the images and build an extensive library of spatio-temporal vehicular density data.

As a first step toward realistic vehicular network modeling, we aim to provide a comprehensive view of the fundamental statistical characteristics of the vehicular traffic density exhibited by the data from four major cities over 45 days. Two main sets of statistical analyses are conducted. The first includes an investigation of the best-fit distribution for the arrival process using various cameras and aggregate city data, while the second is a study of the  long range dependence (LRD) and self-similarity observed in the data. Our early analysis show two main results: i) the empirical distribution of vehicular densities  in most of the cameras and cities follow `log-logistic' and `gamma' distributions. ii)  Consistently, the data showed a high degree of self-similarity over orders of magnitude of time scales, in all cities and for many cameras. This suggests a long-range-dependent process governing the vehicular arrival process in many realistic scenarios. Such result is in sharp contrast to the assumptions of memoryless processes commonly used for vehicular mobility.

The contributions of this work are manifold. (i) To the best of our knowledge, we provide by far the largest and most extensive library of vehicular density data, based on processing of millions of images obtained from ten main cities and thousands of cameras. This addresses a severe shortage of such data sets in the community. The library will be made available to the research community in the future. (ii) We propose a fast algorithm for traffic density estimation to efficiently process millions of image files. (iii) We establish log-logistic and gamma distributions as the most suitable fits for the vehicular density distribution  and provide early evidence of self-similarity exhibited by the traffic at various time scales.

The rest of the document is outlined as follows. Section 2 discusses related work. In Section 3, we discuss our vehicular dataset. In Section 4, we discuss our background subtraction algorithm, and detection and removal of outliers. Statistical analysis of measurements and modeling is illustrated in Section 5. Finally we conclude our paper in Section 6 and give insight into the future work. 
 
\section{Related Work}
Large scale mobility datasets are very important for  the mobile network and computing research community, but collecting them is even more challenging and usually expensive~\cite{Hui:2010:PHM}. In this paper, we propose an inexpensive method to collect  global scale vehicular mobility traces using thousands of freely available  webcams that provide continuous and fine-grained monitoring of the vehicular traffic.

Existing studies in transportation sciences focus on improving road traffic and use of structural engineering methods to resolve issues of congestion, evacuation, and mitigation plans. Initial work\cite{RobertE.Chandler03011958} mainly focused on developing infrastructure for movement of vehicles on roads and bridges. However, in the recent times\cite{halati1997corsim}  much focus has been given to the use of sensor data. The later helps to engineer better traffic conditions, ensuring safety and  management of traffic. For example, inductive loop detectors are equipped to  monitor traffic flows. However, the availability of the data generated from these sensors is not readily available to the general public. Second, studies\cite{bychkovsky2006measurement} do not necessarily focus on vehicular networks, traffic modeling, and characterization.  In spite of data availability problems, surprisingly there is a large deployment of publicly available online web cameras, which can be used to  monitoring and modeling traffic. In our work, we take advantage of these free webcams. To our knowledge\textit{ we are the first to identify the power and usability of these free web cameras for the purpose of modeling and characterizing the traffic across globe}.   

Simulation tools like CORSIM\cite{halati1997corsim} and VISSIM\cite{Lownes:2006:VMS:1218112.1218368} are geared to model specific scenarios for planning future traffic conditions on a micro-mobility and small scale level. In this work, we focus on the aspect of macro-mobility to model vehicular movements in form of flow densities to analyze traffic on huge scale. From a networking perspective, mobility models\cite{bychkovsky2006measurement,1354509} and routing \cite{Zhang:2007:SBD:1287853.1287876} techniques investigate how mobility   impact the performance of routing protocols \cite{DBLP:conf/infocom/BaiSH03}. If the mobility model is unrealistic then routing performance is questionable. So, we need models inspired from real data sets. By way of this work, we believe a comprehensive set of parameters can be extracted to develop such models. 


In a recent work,  Bai et. al \cite{Bai:2009:SVV:1614269.1614278} analyzed spatio-temporal variations in vehicular traffic from the purpose of inter-vehicle communications. Data collected from realistic scenarios shows the effectiveness of  exponential model for highway vehicle traffic. On the same line, quantitative characteristics of vehicle arrival pattern on highways  is studied in \cite{meng:864}. By using real highway traffic data, the study examines the existence of self-similarity characteristics on vehicle arrival data and finds that time headway of vehicles on the highways follows the heavy-tailed distribution. These findings enrich traffic modeling, but carried out on very small sample of data and mainly localized to one or two locations. In our study, we use 45 days of vehicular imagery data from four cities to model traffic and characterize the density distribution.

A principle activity related to our work is image processing and efficient retrieval of traffic  information from these images. Many studies{\cite{RobertE.Chandler03011958} have been carried out that look into aspects of  both background subtraction\cite{1400815,784637} and object detection\cite{1038171}.  In former methods\cite{springerlink:elgammal}, difference in the current and reference frame is used to identify objects. In  detection approaches\cite{10.1109/TPAMI.2006.104},  learning the object features (shape, size etc.) are used to detect and classify them. In our work, we are using a temporal methods for background subtraction to calculate a relative numerical value instead of counting cars. In our work we find background subtraction is much faster than object detection, which is discussed in detail in later section.

\begin{table*} 
\centering
\caption{Global Webcam Datasets}
\scalebox{1}{
\begin{tabular}{|c| c| c| c | c | c |} \hline
\textbf{City} & \textbf{ \# of Cameras} & \textbf{Duration} & \textbf{Interval} & \textbf{  Records} & \textbf{Database Size} \\ \hline
Bangalore & 160 & 30/Nov/10 - 01/Mar/11 & 180 sec & 2.8 million & 357 GB  \\\hline
Beaufort & 70 & 30/Nov/10 - 01/Mar/11 & 30 sec. & 24.2 million & 1150 GB \\  \hline
Connecticut & 120 & 21/Nov/10- 20/Jan/11 &  20 sec. & 7.2 million & 435 GB \\ \hline
Georgia & 777 & 30/Nov/10 - 02/Feb/11 & 60 sec. & 32 million & 1400 GB \\  \hline
London & 182 & 11/Oct/10 - 22/Nov/10 & 60 sec. & 1 million &  201 GB \\ \hline
London(BBC) & 723 & 30/Nov/10 - 01/Mar/11 & 60 sec. & 20 million & 1050 GB \\ \hline
New york & 160 & 20/Oct/10 - 13/Jan/11 & 15 sec. & 26 million & 1200 GB \\  \hline
Seattle & 121 & 30/Nov/10 - 01/Mar/11 & 60 sec. & 8.2 million & 600 GB \\  \hline
Sydney & 67 & 11/Oct/10 - 05/Dec/10 & 30 sec. &2.0 million &  350 GB \\  \hline
Toronto & 89 & 21/Nov/10 - 20/Jan/11 & 30 sec. & 1.8 million & 325 GB \\  \hline
Washington & 240 & 30/Nov/10 - 01/Mar/11 & 60 sec. & 5 million & 400 GB \\\hline
\textbf{Total}	&  \textbf{2709} &\textbf{-} &\textbf{-} & \textbf{125.2 million} & \textbf{7468 GB} \\
\hline\end{tabular}}
\label{dataset}
\end{table*}

\begin{figure}
\centering
\epsfig{file=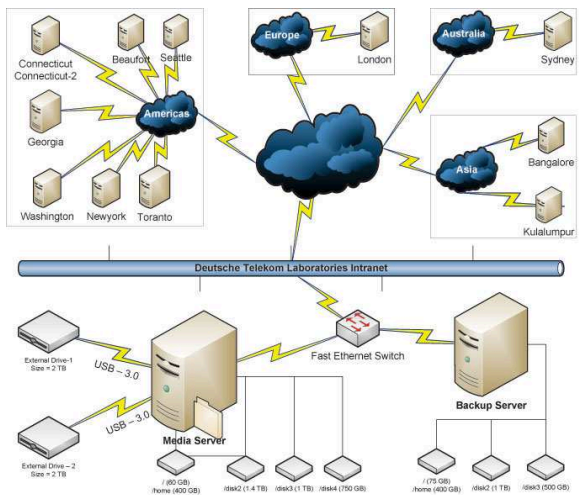,width=3in}
\caption{Infrastructure for measurement collection}
\label{infsetup}
\end{figure}

\begin{figure}[!t]
\centering
\mbox{\subfigure[London]{
   \includegraphics [ scale=.21]{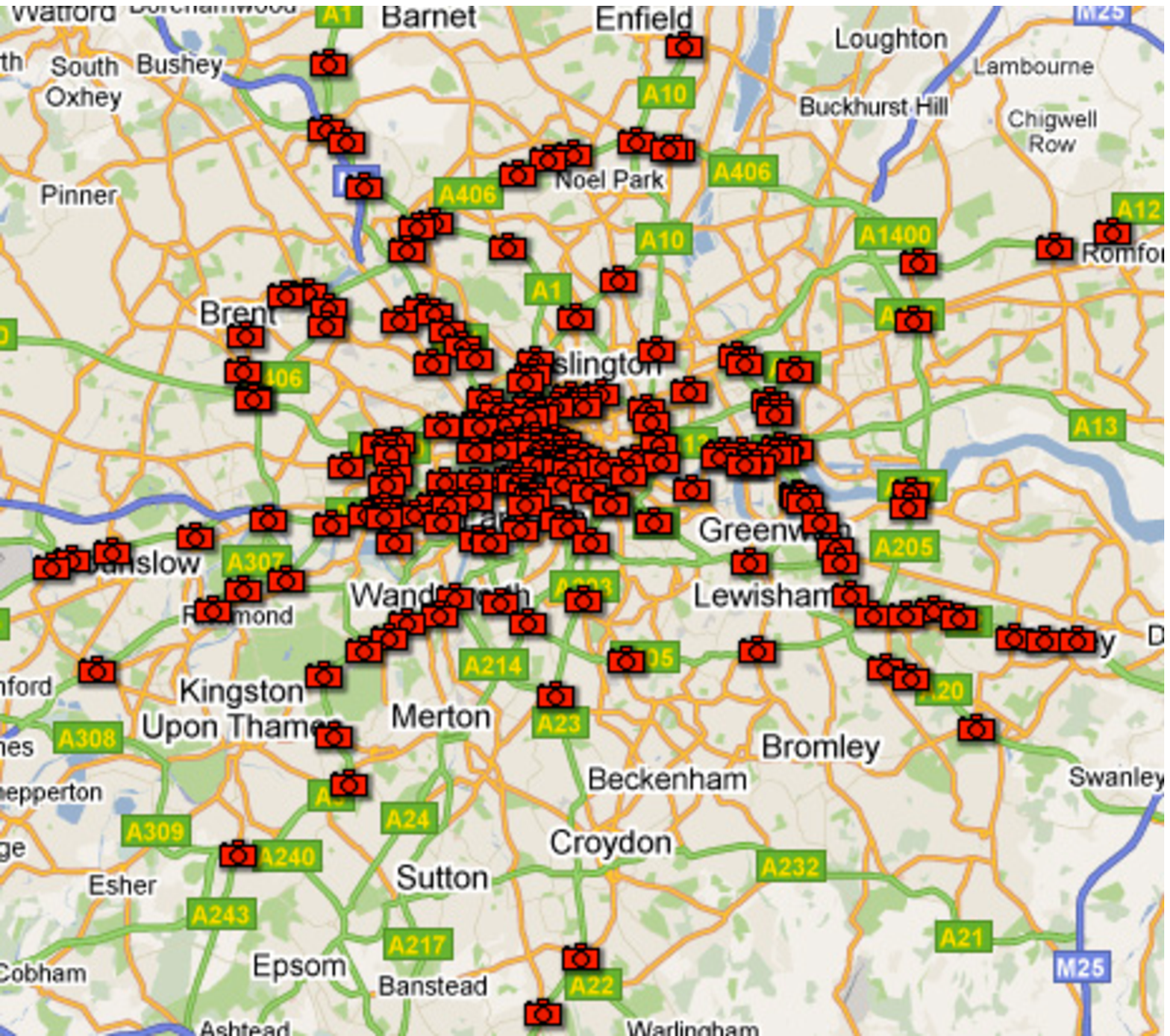}
   \label{fig:subfig11}
 }
\quad
\subfigure[Sydney]{
   \includegraphics[ scale=0.21] {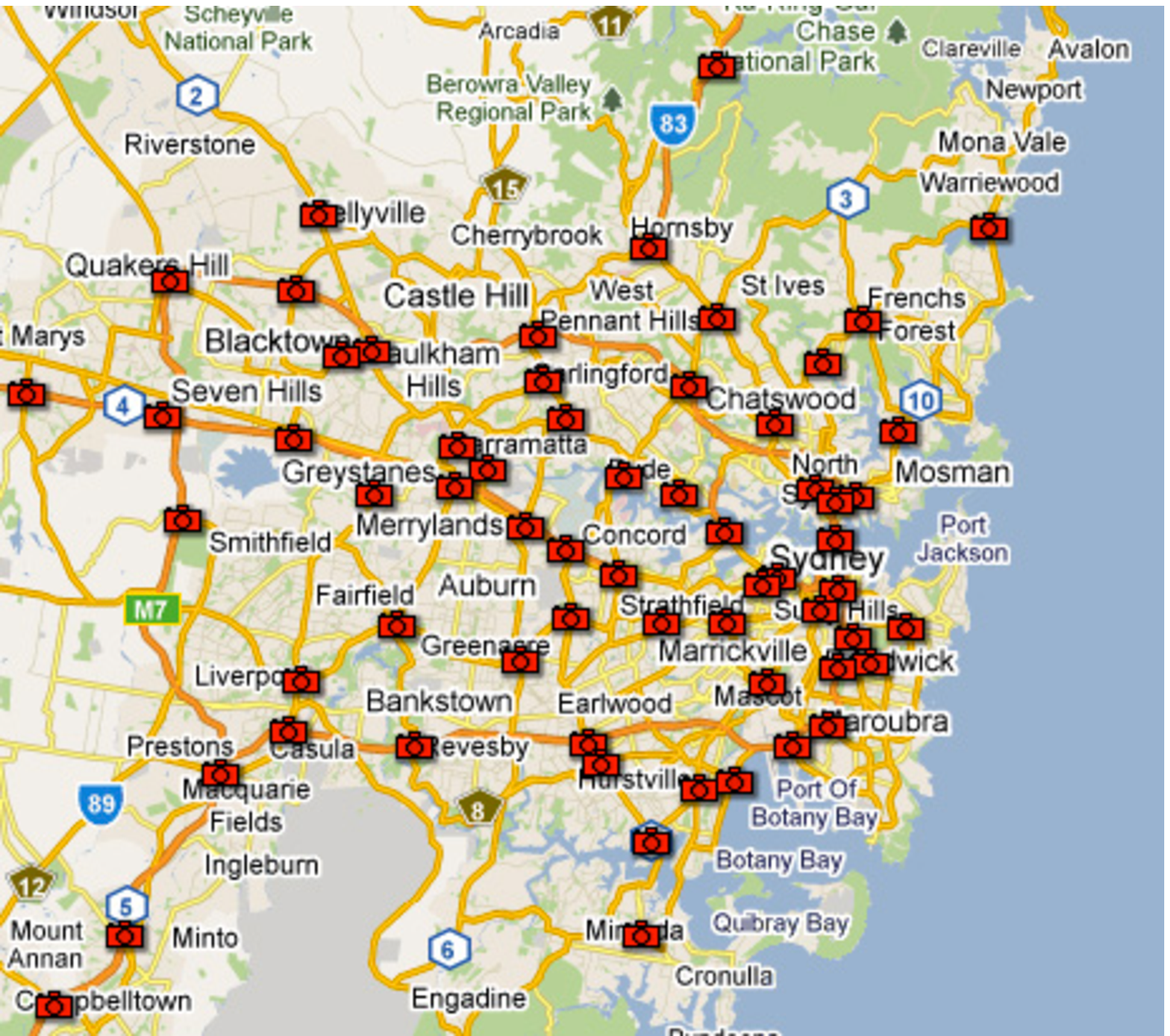}
   \label{fig:subfig12}
 }
  }
\label{citymaps}
\caption{\small{Traffic cameras in London and Sydney. The red dots show the location of cameras deployed.}}
\end{figure}

\newpage

\section{Data Collection}
There are thousands, if not millions, of outdoor cameras currently connected to the
Internet, which are placed by governments, companies, conservation societies, national
parks, universities, and private citizens.                                                                            Outdoor
webcams are usually mounted on a roadside pole with easy accessibility, installation
and maintenance, and they have seen enormous applications not only in adaptive traffic
control and information systems, but also in monitoring the weather conditions,
advertising the beauty of a particular beach or mountain, or providing a view of animal
or plant life at a particular location.  We view the connected global network
of webcams as a highly versatile platform, enabling an untapped potential to monitor
global trends, or changes, in the flow of the city, and providing large-scale data to realistically model vehicular, or even human, mobility.

In this section, we  introduce the methodology for the data collection and give a high level statistics of the data traces.    We collect vehicular mobility traces using the online webcam crawled by our crawler. A majority of these webcams are deployed by the Department of Transportations (DoT) in each city. They are used to provide real time  information about road traffic conditions to general public via online traffic web cameras. These web cameras are basically installed on  traffic signal poles facing towards the roads of some prominent intersections throughout city and highways. At regular interval of time, these camera captures  still pictures of on-going road traffic and send them in form of feeds to the DoTs media server. For the purpose of this study, we chose 10 cities with large number of webcam coverage and took the permission from concerned DoTs to collect these vehicular imagery data for several months. We cover cities in North America, Europe, Asia, and Australia. In Fig.-\ref{infsetup}, we show our experimental infrastructure to download and maintain the image data. Since these cameras provide better imagery during the daytime, we limit our study to download and analyze them only during such hours. On average, we download 15 Gigabytes of imagery data per day from over 4700 traffic web cameras, with a overall dataset of 6.5 Terabytes and containing around 120 millions images. Table-\ref{dataset} shows the high level statistics of datasets we collected. Each city has a different number of deployed cameras and a different interval time to  capture images. For example, cameras for the city of Sydney capture images at an interval of one minute  while for the state of connecticut the interval time between two consecutive snapshots is only 20 seconds. The wide spread geographical deployment of these cameras covering major sections of city and highways. Fig.-\ref{citymaps} give an example of the camera deployments in the city of London and Sydney by mapping the Global Positioning System (GPS) location of the cameras to Google maps. The area covered by the cameras in London is $950km^2$ and that in Sydney is $1500km^2$. Hence, we believe our study will be comprehensive and will reflect major trends in traffic movement of cities.        

\section{Algorithm to Extract Traffic Densities}
We aim to estimate traffic density on roads considering the number of vehicles or pedestrians crossing the road. We have a sequence of images $(I_1(x,y) + I_2(x,y)... +I_z(x,y))$ captured by webcams. Considering our problem, we have to be able to separate information we need, e.g. number of vehicles and pedestrians from the back ground image which is normally road and buildings around. The main factor that can distinguish between vehicles and background image (road, buildings) is the fact that the vehicles are not in a stationary situation for a long period of time, however the back ground is stationary.  The solution for the problem then seems to be applying a sort of high pass filtering over a sequence of images captured by a webcam over time. The high pass filter removes the stationary part of the images (road, buildings, etc.), and keeps the moving components (mainly vehicles). In order to implement such a high pass filter, we subtract result of a low pass filter over a sequence of images, from each still image. This is practically equal to implementing a high pass filter over sequence of images. In order to obtain low pass filtering effect, we run a moving average filter over a time sequence of images obtained from one webcam. The duration of moving average filter can be adjusted in an adhoc way. The moving average filter is simply implemented by averaging over intensity map for several images in a certain duration. At the output of moving average filter, the intensity of each pixel is obtained by averaging intensity of corresponding pixels in the interval.  The output of the moving average filter (low pass filter) is normally the required background image, which is still image of street and buildings. Therefore, subtracting each image from the output of low pass filter, gives us the moving components (e.g. vehicles). This is in fact the high pass component of the image over time.

Having the high pass component of the image, the vehicles are highlighted from background. One may then use regular object detection techniques to identify and count number of vehicles in the high pass filtered image. However, applying such techniques may require heavy load of computation, and in the same time it can be unnecessary. As an alternative, we  simply counting number of active pixels (pixels with a value higher than a certain threshold). Such a process can be much faster than detecting and counting objects in an image. In the same time, it can be much more effective, because we are looking for the percentage of the street (road) which is covered by vehicles (as an indicator of how crowded is the street), rather than number of vehicles. Number of vehicles can not be necessarily a good indicator of crowdedness, as a long vehicle may introduce more traffic than a small one. Secondly, it overcomes the issues that object detection algorithm face in conditions of severe congestions. One of them is visibility of boundary  contours used to separate objects from one another.    In contrary, counting number of active pixels can indicate what percentage of the road is covered, no matter how many vehicles are in the road.

Said that, consider an image can be represented as

\begin{center}
$I(x,y) = L(x,y) + T(x,y)  + N(x,y) $
\end{center}

where $I(x,y)$ is the captured image, $L(x,y)$ is our low pass filter and $T(x,y)$ and $N(x,y)$ are respectively the traffic and associated noise with the images. In first step, we generate a low pass filter using the aforementioned technique of moving average. Initially, we average  a give data pixel with its right and left neighbors. For the purpose of this study, we kept the number of its neighbors $z=100$. The averaging results in the removal of dominant trends. These dominant trends are $T(x,y)$ and $N(x,y)$.  This low pass filter remains constant for one camera,

\begin{center}
$L(x,y) = (I_1(x,y) + I_2(x,y)... +I_z(x,y))/z$   
\end{center}

To get the traffic density associated with an image we subtract the low pass filter and set a threshold ($\tau$) to reject a resulted pixel value below it so as to reduce the effect of noise (shadows etc.) $N(x,y)$. In summary,

\begin{center}
$I^{'}(x,y) =I(x,y)  - L(x,y) $
\end{center}

Such that $I^{'}(x,y) > \tau$. Later, we convert the image to grayscale $I^{''}(x,y)$ and sum the pixels to get the traffic density ($d$).

\begin{center}
$d=\displaystyle\sum_{x=0}^{m}\sum_{y=0}^{n}I^{''}(x,y)$
\end{center}

 \begin{figure}
\centering
\mbox{
\subfigure[Outliers Present]{
   \includegraphics [ width=1.51in]{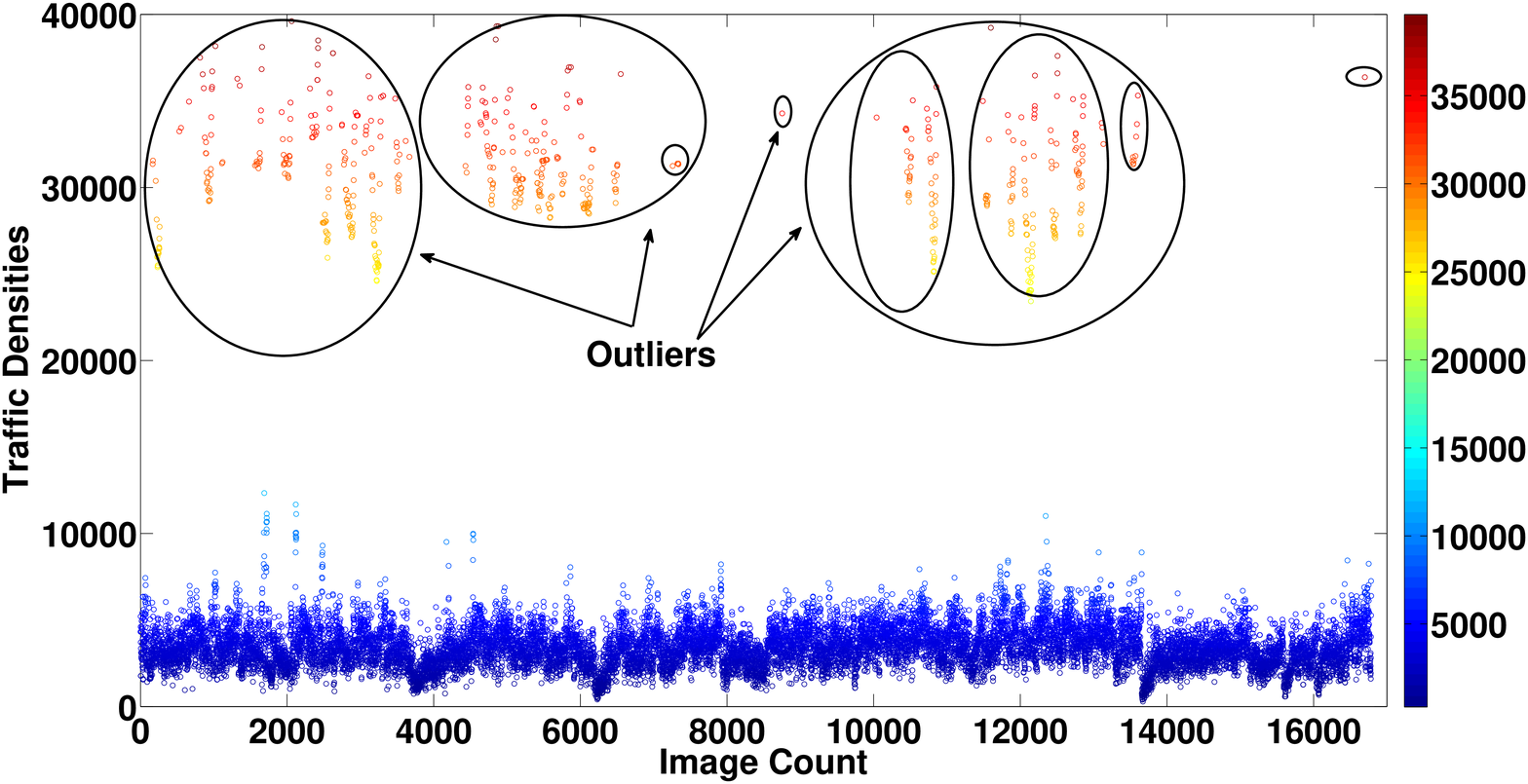}
 }
\subfigure[Outliers Removed]{
   \includegraphics[width=1.51in] {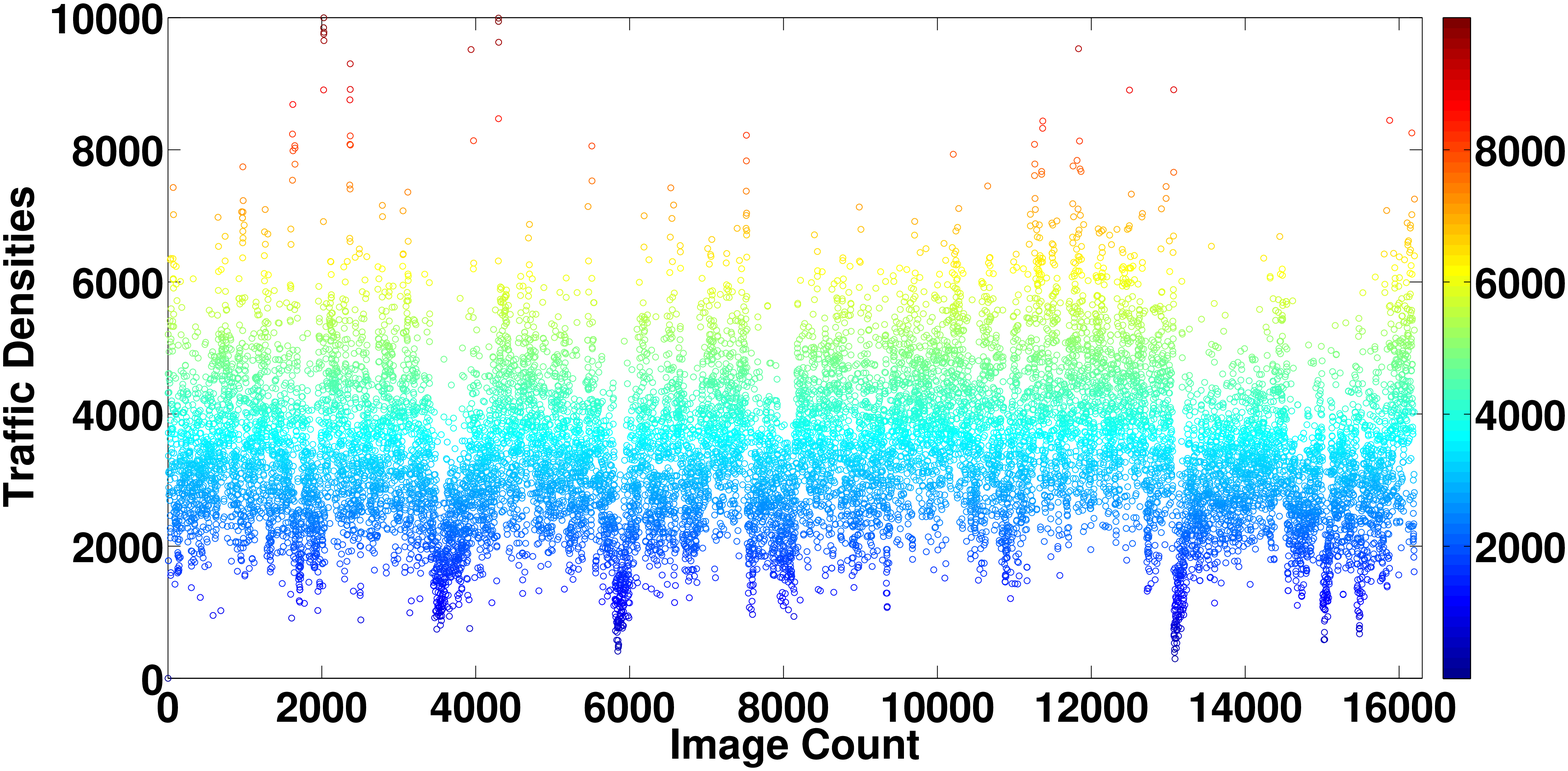}
}
}
\label{outliers}
\caption{Outliers detection and removal. (a) Outliers detection  by encircling them (b) Factual traffic  density distribution.}
\end{figure}

\subsubsection*{Outliers Detection and Removal}
An important aspect of collecting images on such a large scale requires automated processes to manage and extract useful information.  As mentioned, different cameras have different refreshing rate, we have to continuously download images at a specific time-interval for each camera. To ensure that we are not missing even a single traffic snapshot, we keep our download time-interval a little shorter than the camera refreshing rate. However, this results in few duplicate images that we filter out as a first step towards outliers detection and removal. Normally, the downloaded data set contain images, which are the snapshot of vehicular traffic on the roads. But in many instances, the images are corrupted with zero sized or with extraneous bytes (noise). Next, if the camera instrument is non-functional or has mechanical errors, the traffic monitoring server replaces current traffic snapshot with error notification image.

 The challenge here is to detect all such errors and remove them before modeling and statistical analysis.  The analysis become more complex as we do not know the kind of distribution underlying and hence any statistical techniques that rely on some distribution (boxplot etc) cannot be used. We used semi-supervised learning and data mining to overcome the challenges of outliers detection and removal in millions of traffic images.

 In our case, we treat data set $X$  containing all types of images as $X=\{x_i,x_2,x_3,...,x_n\}$. Later on we divide this set into two parts: the data points in $X_l=\{x_1,x_2,x_3,...,x_l\}$ mapped to labels in $Y_l=\{y_1,y_2,y_3,...,y_{l}\}$. The provided input features includes but not limited to image size, color depths, multi-channel color arrays and image segmentation stderrs for detecting outliers. The second part contains points with unknown labels represented as 

\begin{center}
$X_u=\{x_{l+1},x_{l+2},x_{l+3},...,x_{l+u}\}$ 

\end{center}
such that $u>>l$. The already known and learned labeled point are later used to find cluster boundaries and assigning class to each cluster.

In this case, we used low density separation assumption that help to cut the dataset into clusters. The identified clusters are separated out as outliers, which are mostly distant from the regular traffic density data. In Fig-3, we compare the results of detecting and removing the outliers.\\       
\begin{figure}
\centering
\mbox{
\subfigure[$d=2023,0.28$]{
   \includegraphics [scale=.2]{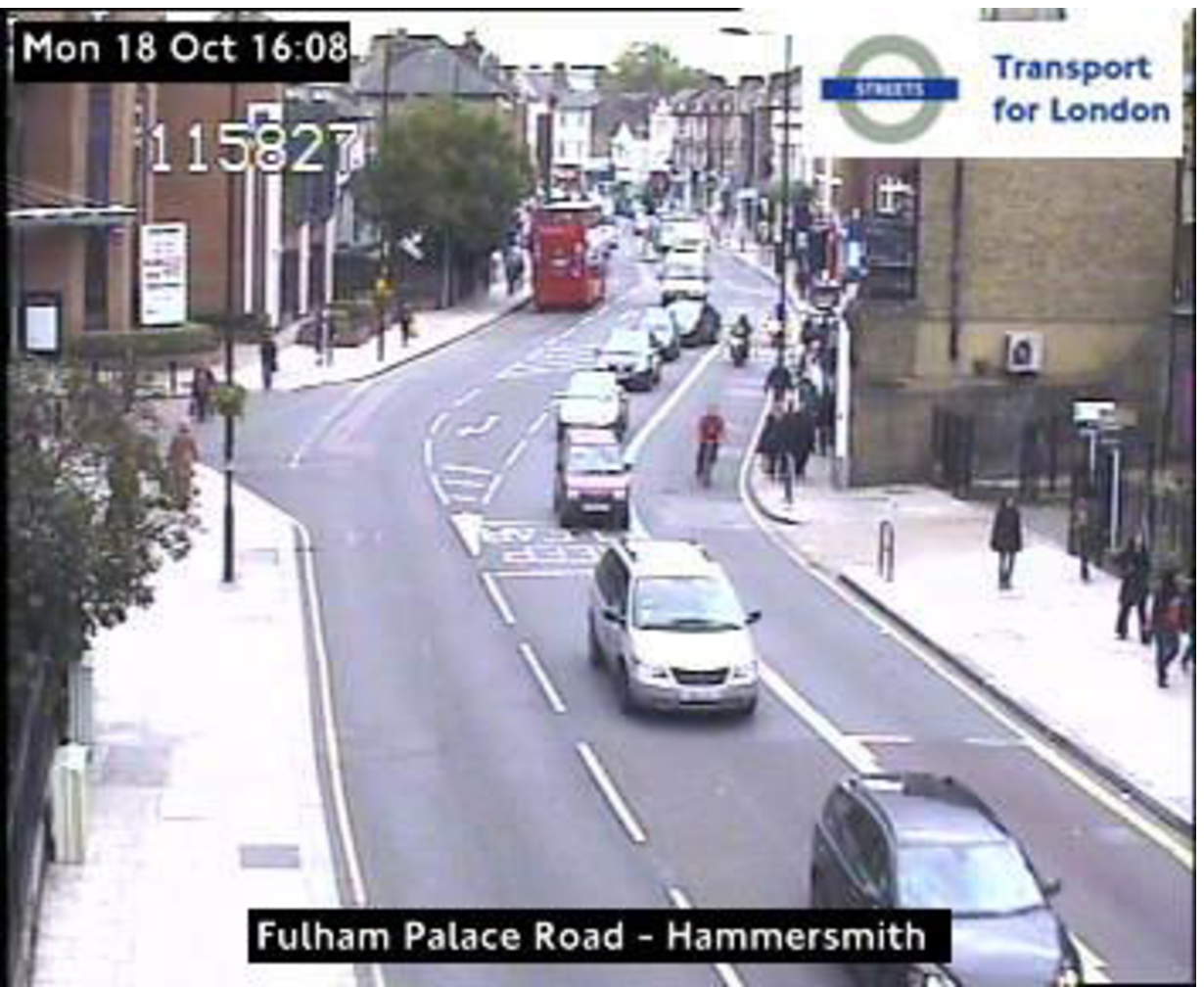}
 }
\subfigure[$d=5400,0.55$]{
   \includegraphics[scale=.2] {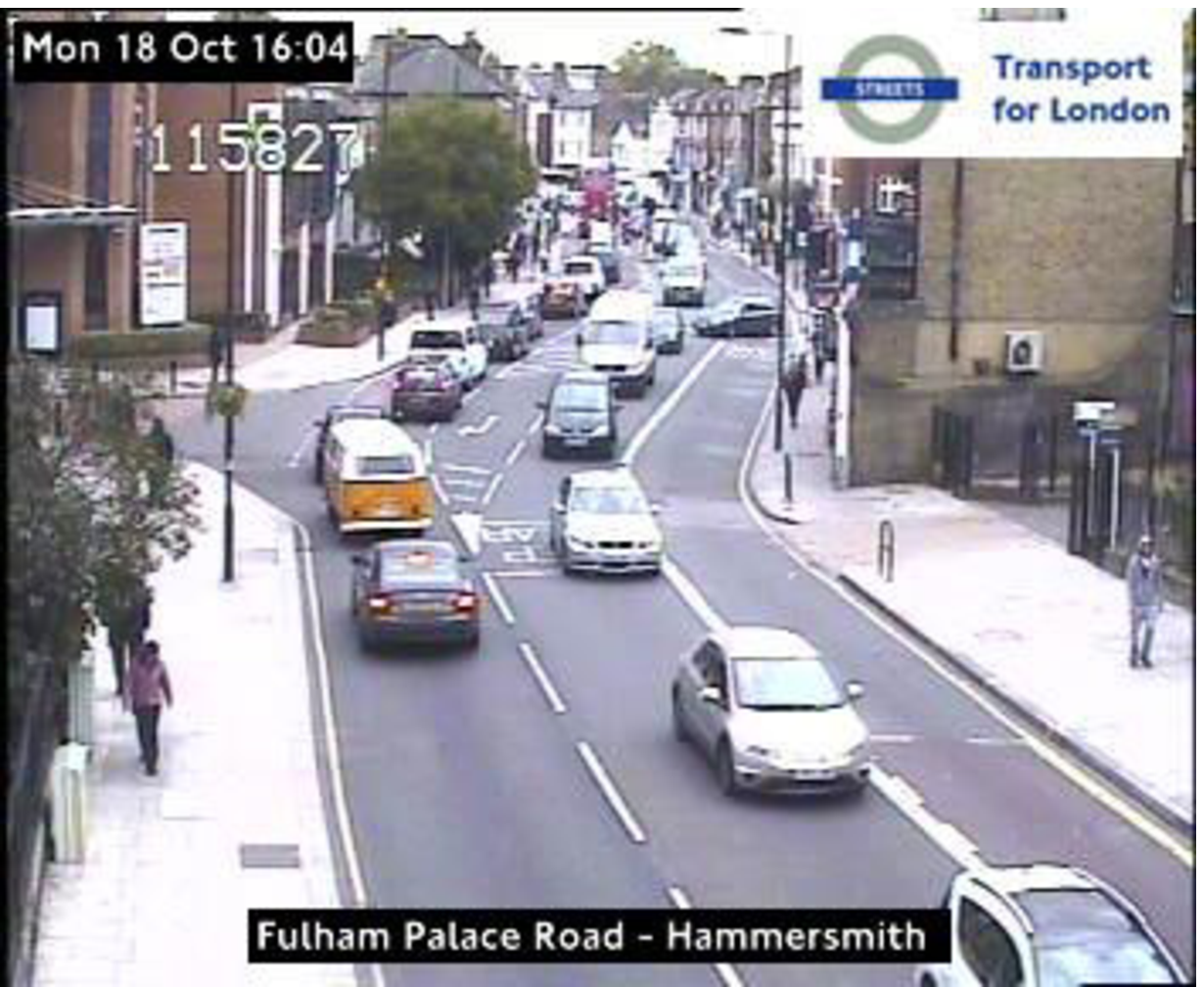}
}
\subfigure[$d=9230,0.93$]{
   \includegraphics[scale=.2] {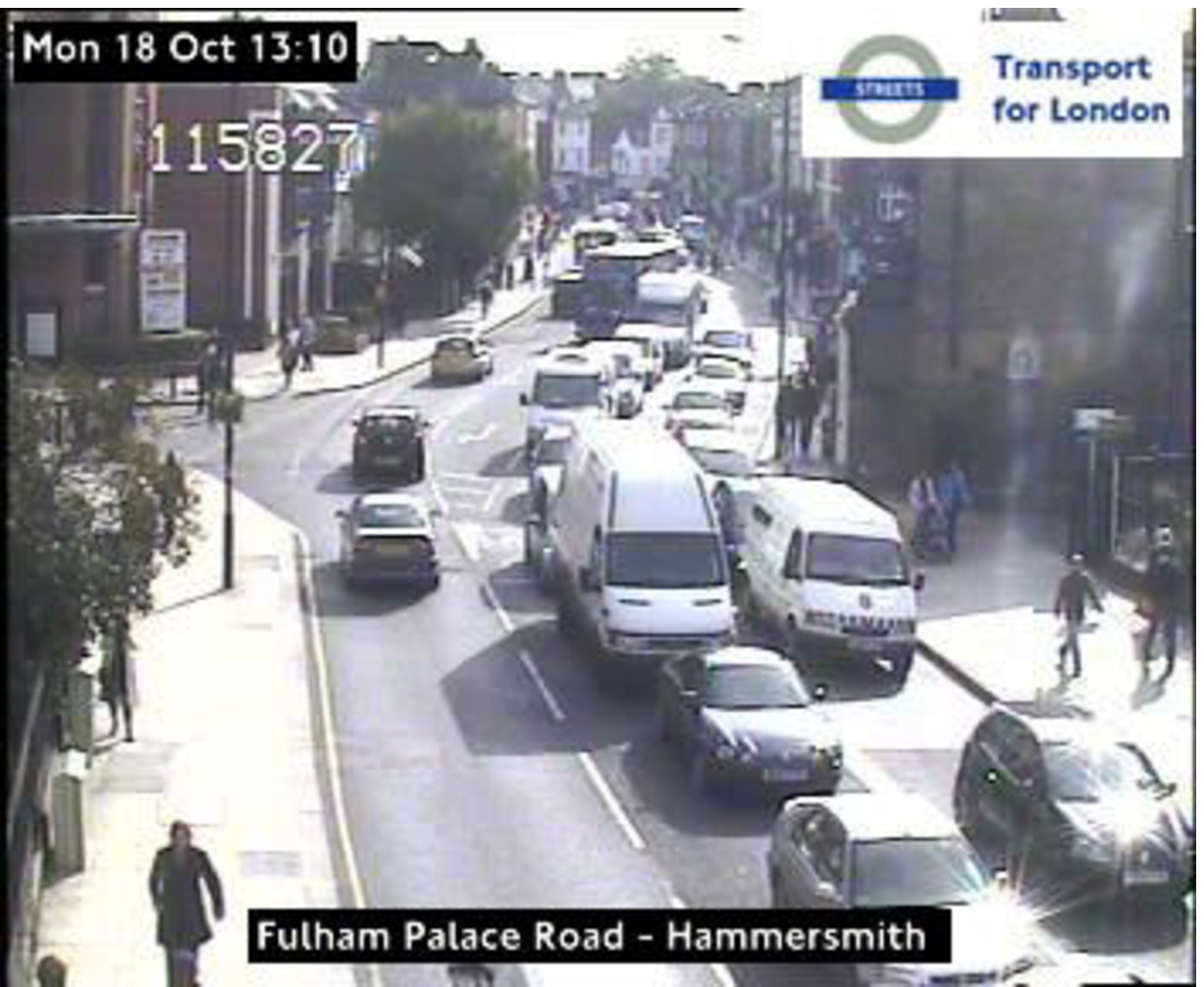}
}
}
\label{figs3}
\caption{A series of pictures for same intersection but varying [(a)low/(b)medium/(c)high] traffic intensities. This variation is captured by density  parameter $d$. The first values is the result of background subtraction and later is the normalized value. }
\end{figure}

 \begin{figure}[!ht]
\centering
\epsfig{file=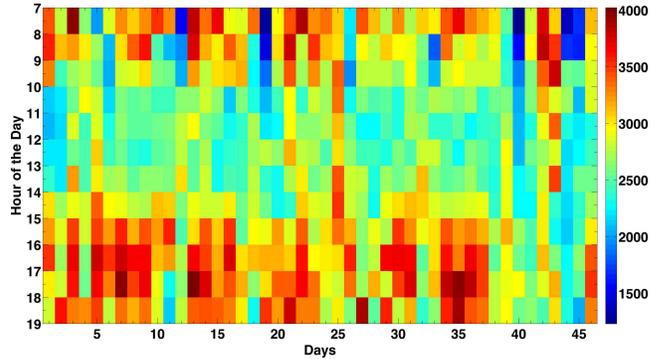,width=3.33in}
\caption{Traffic arrival process on hourly basis for 45 days. A regular pattern of high traffic intensity during morning and evening hours is evident. }
\label{imm}
\end{figure}

 \begin{figure*}
\centering
\mbox{
\subfigure[Connecticut]{
   \includegraphics [scale=0.2]{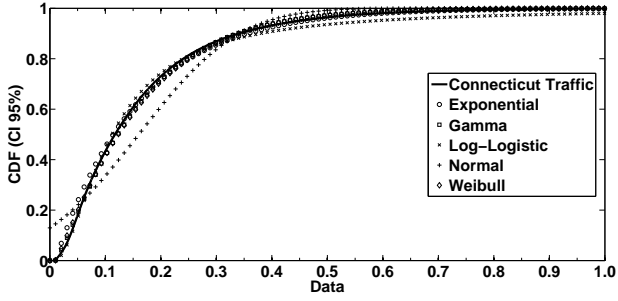}
   \label{fig:c1}
 }
\quad
\subfigure[London]{
   \includegraphics[ scale=0.2] {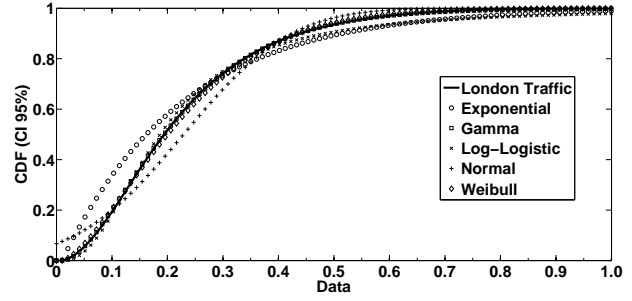}
   \label{fig:c2}
 }
}
\mbox{
\subfigure[Sydney]{
   \includegraphics [ scale=0.2]{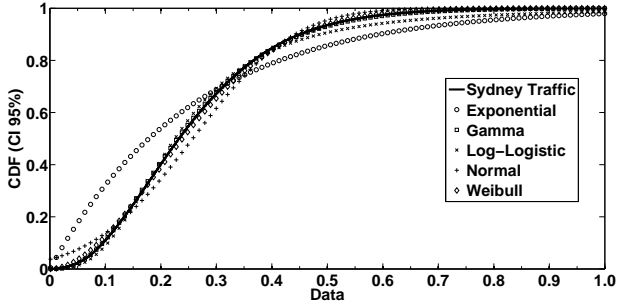}
   \label{fig:c3}
 }
\quad
\subfigure[Toronto]{
   \includegraphics[ scale=0.2] {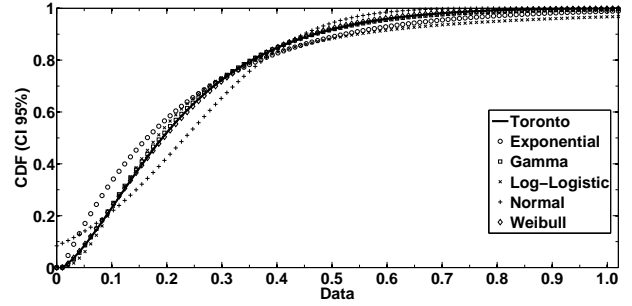}
   \label{fig:c4}
 }
}
\caption{Modeling the distribution for aggregate traffic densities. }
\label{cpdf}
\end{figure*}

\begin{sidewaysfigure*}
\centering
\mbox{
\subfigure[Connecticut(L)]{
   \includegraphics [scale=.17]{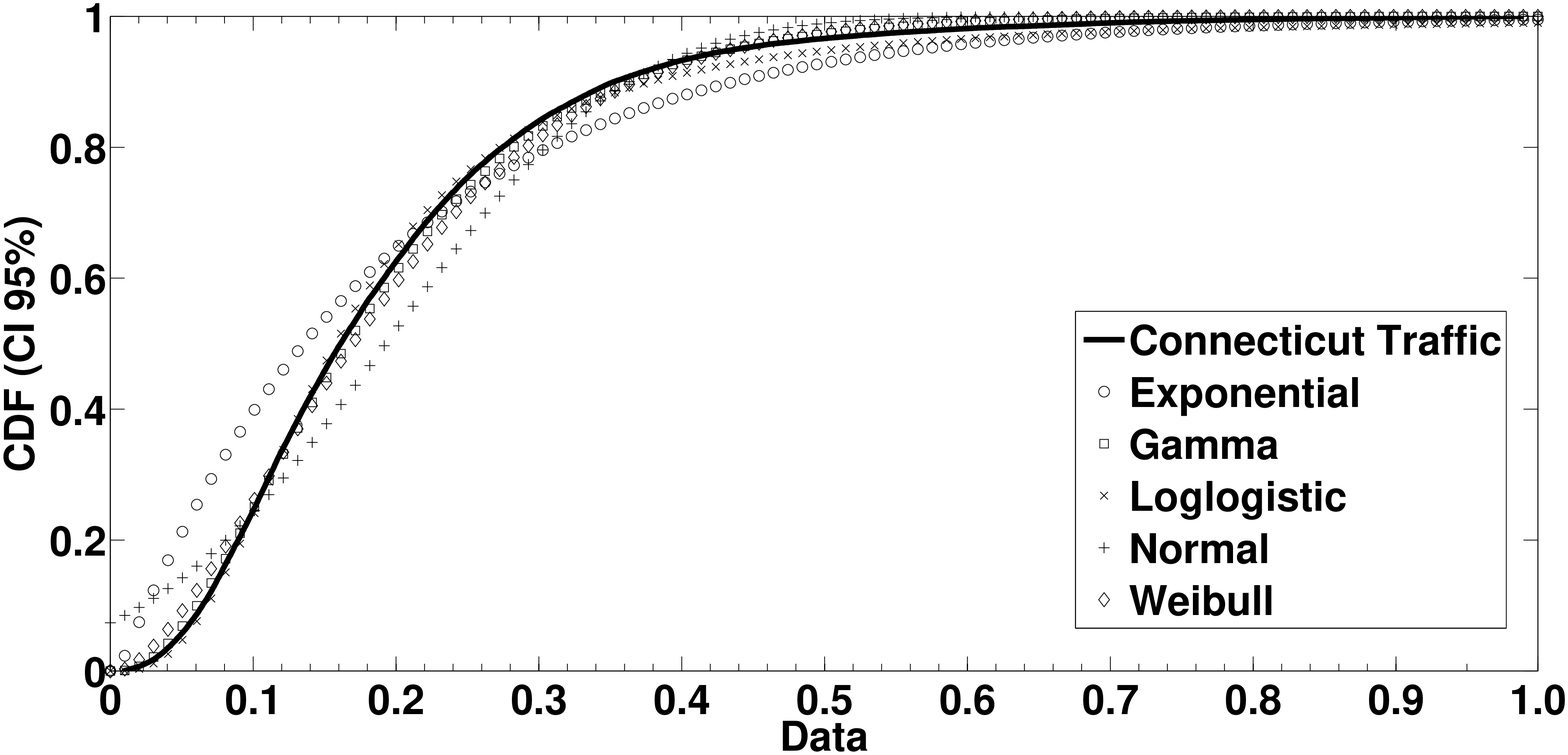}
   \label{fig:cc1}
 }
 \subfigure[Connecticut(M)]{
   \includegraphics [scale=.17]{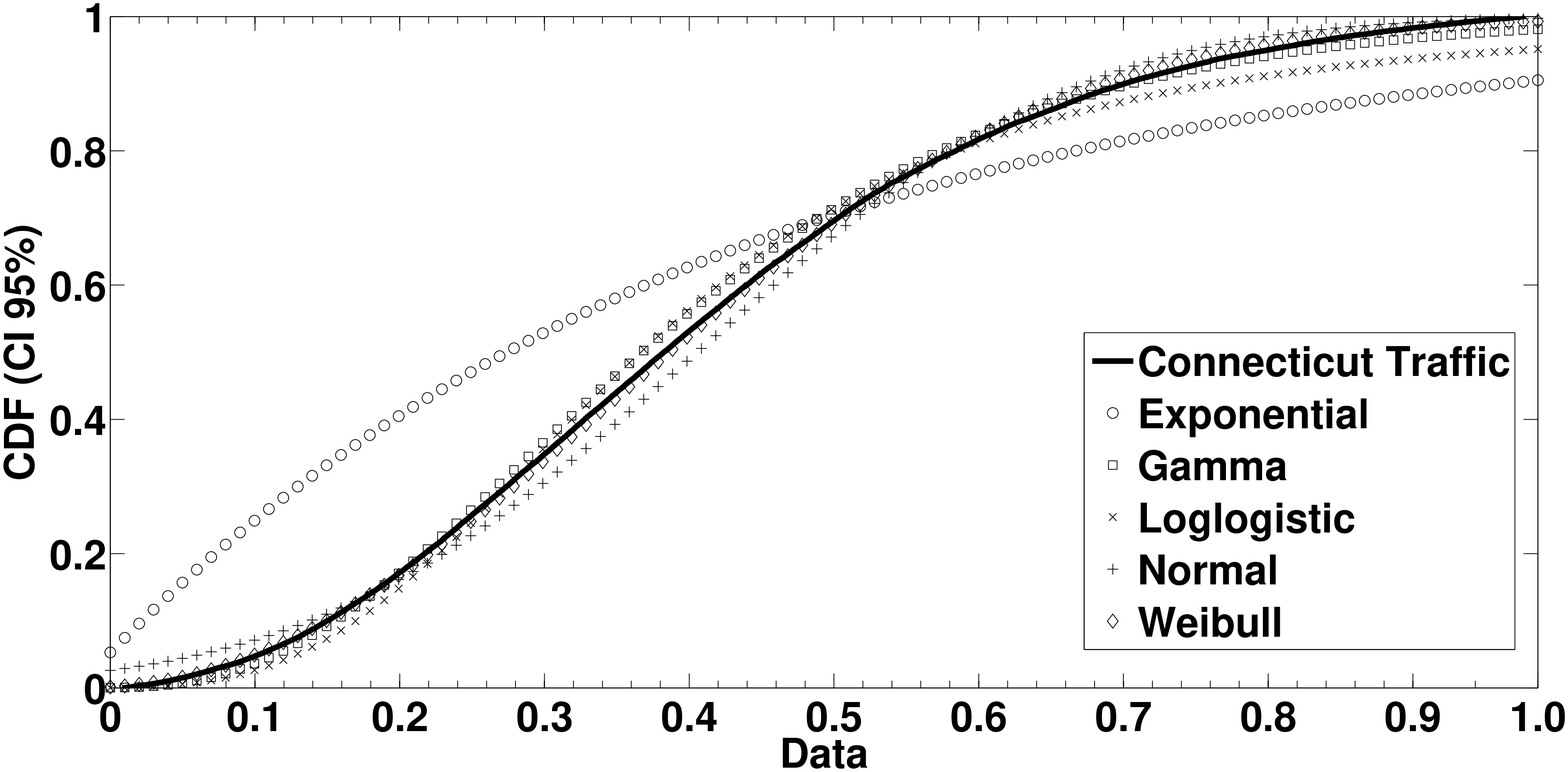}
   \label{fig:cc2}
 }
 \subfigure[Connecticut(H)]{
   \includegraphics [scale=.17]{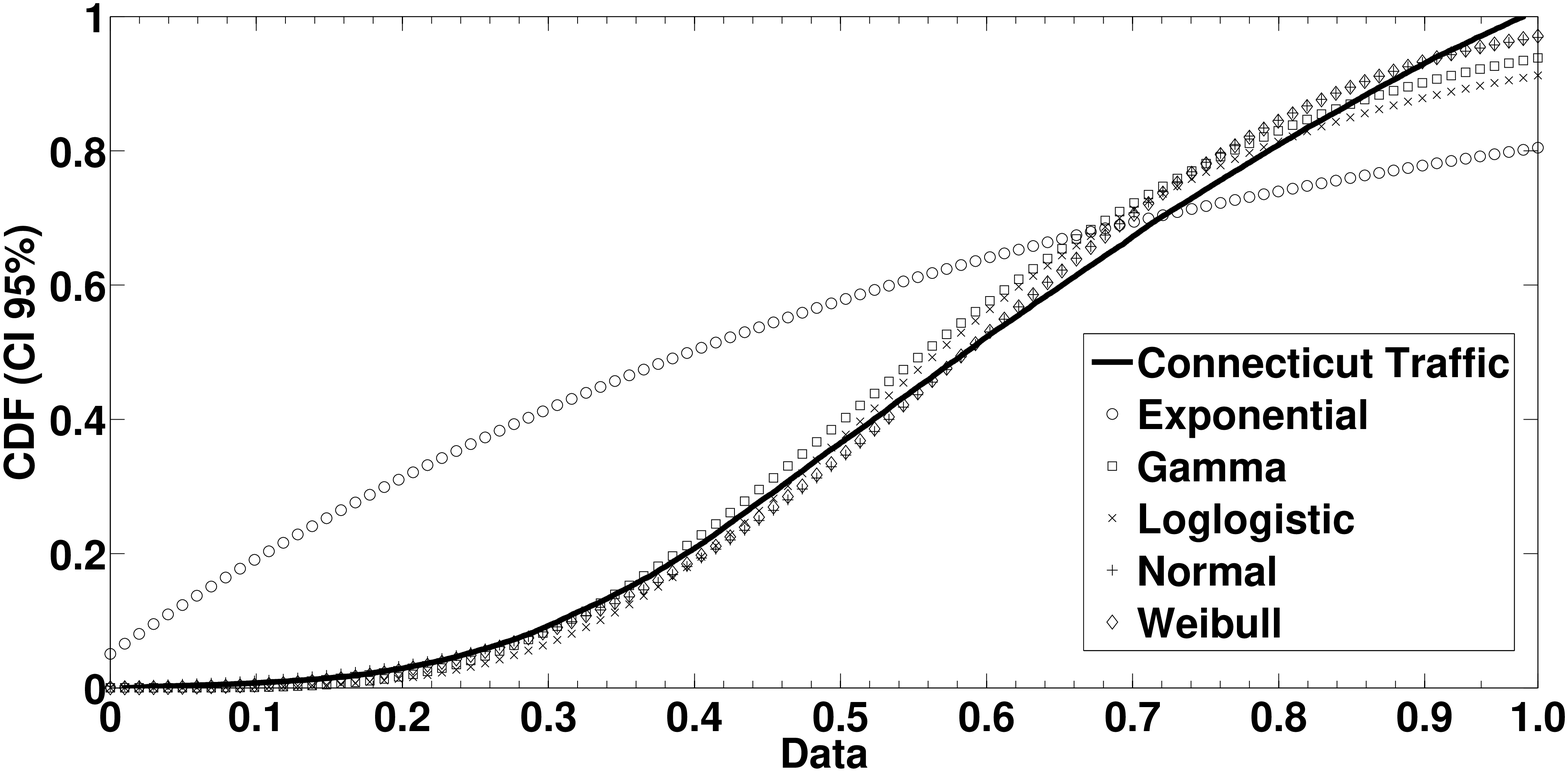}
   \label{fig:cc3}
 }
 }
 \mbox{
 \subfigure[London(L)]{
   \includegraphics [scale=.17]{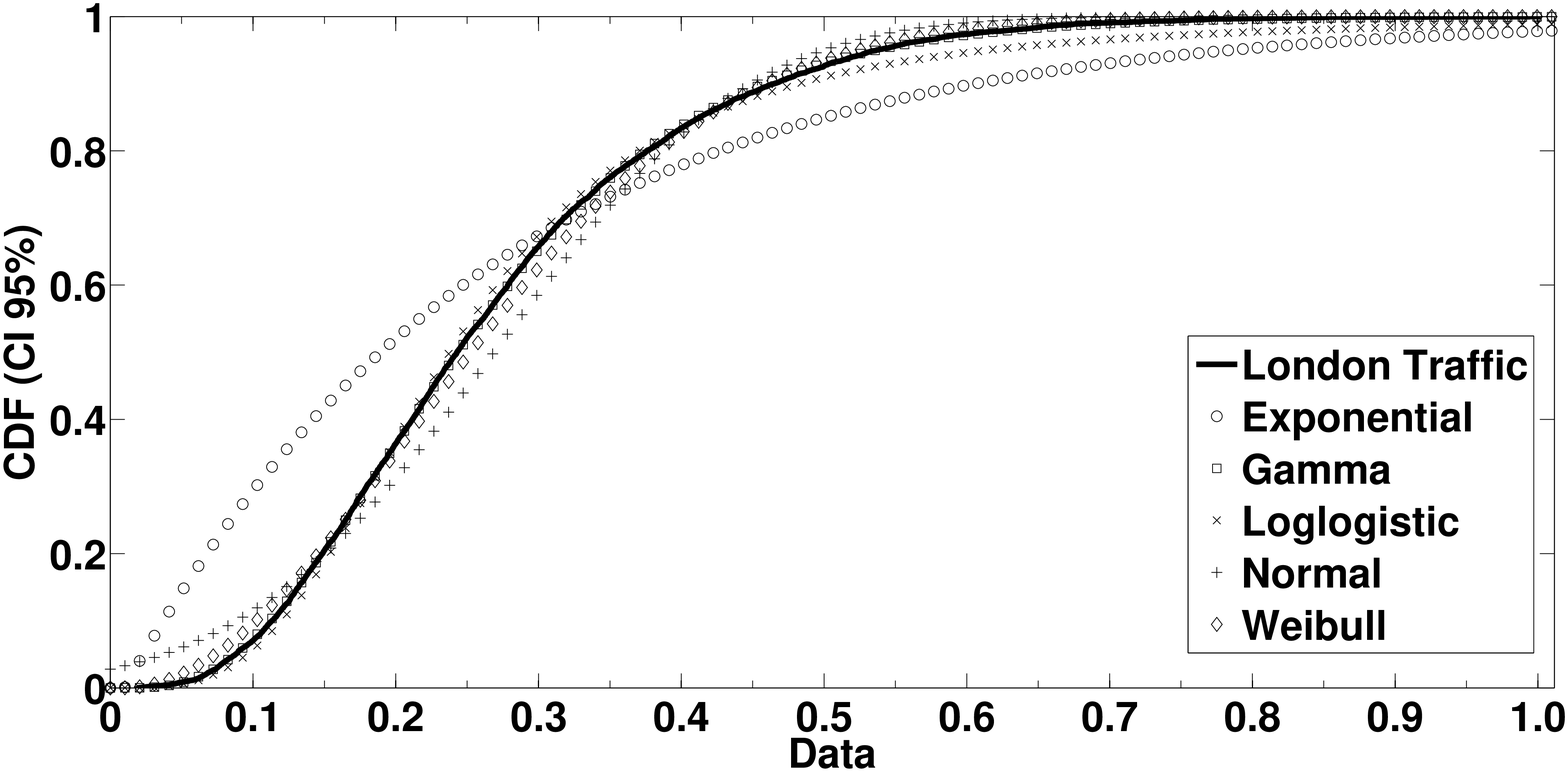}
   \label{fig:l1}
 }
 \subfigure[London(M)]{
   \includegraphics [scale=.17]{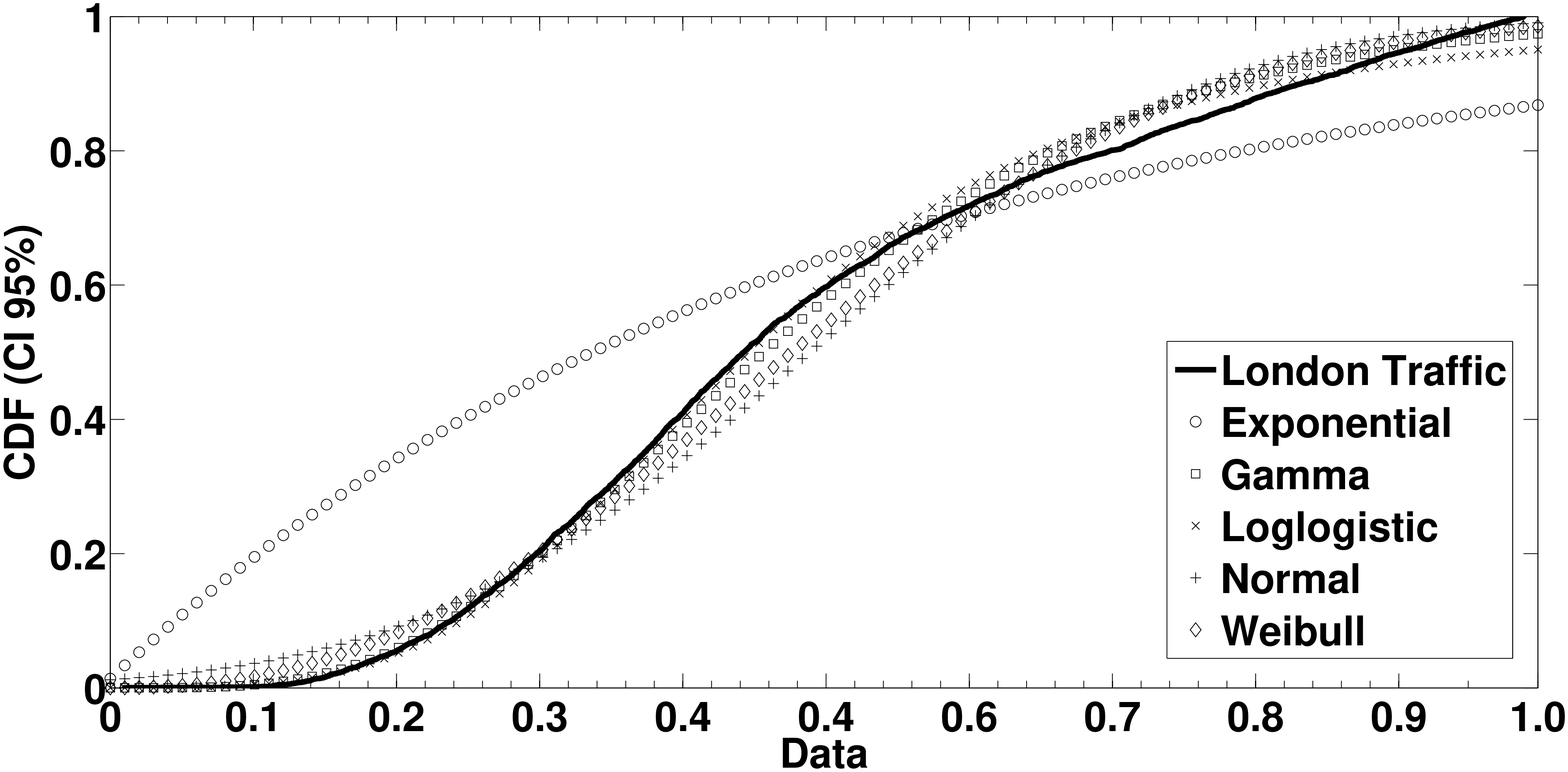}
   \label{fig:l2}
 }
 \subfigure[London(H)]{
   \includegraphics [scale=.17]{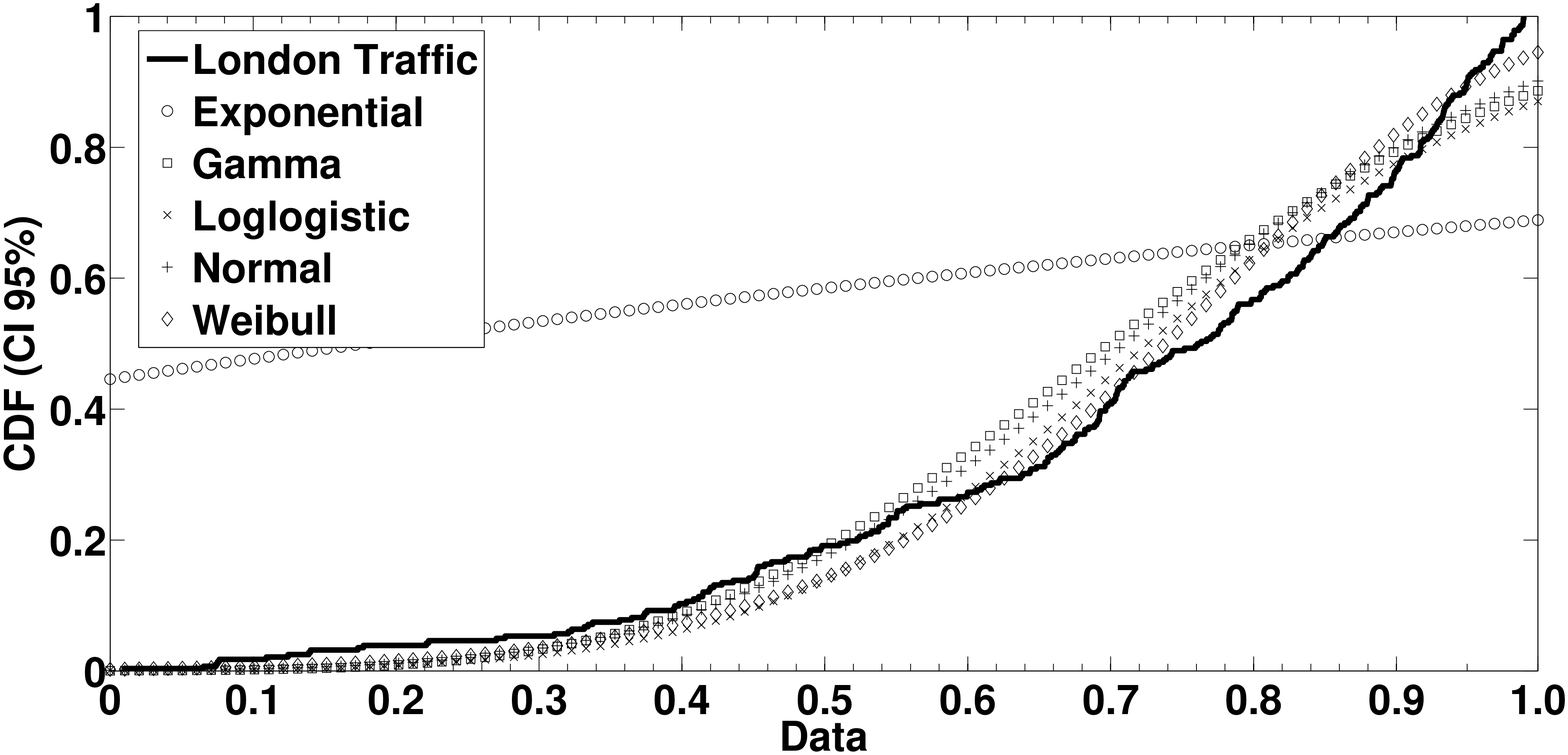}
   \label{fig:l3}
 }
}
\mbox{
\subfigure[Sydney(L)]{
   \includegraphics [scale=.17]{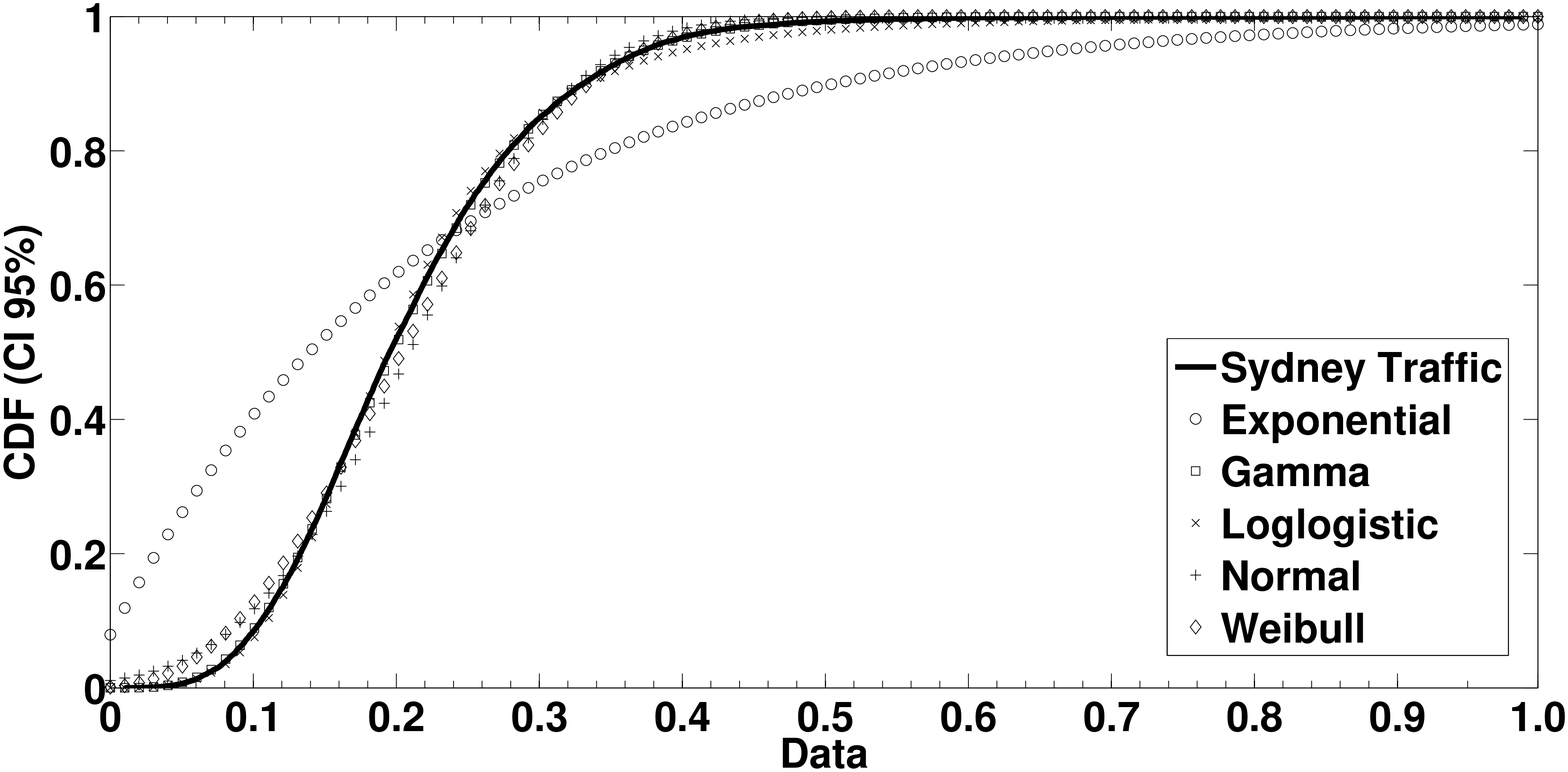}
   \label{fig:s1}
 }
 \subfigure[Sydney(M)]{
   \includegraphics [scale=.17]{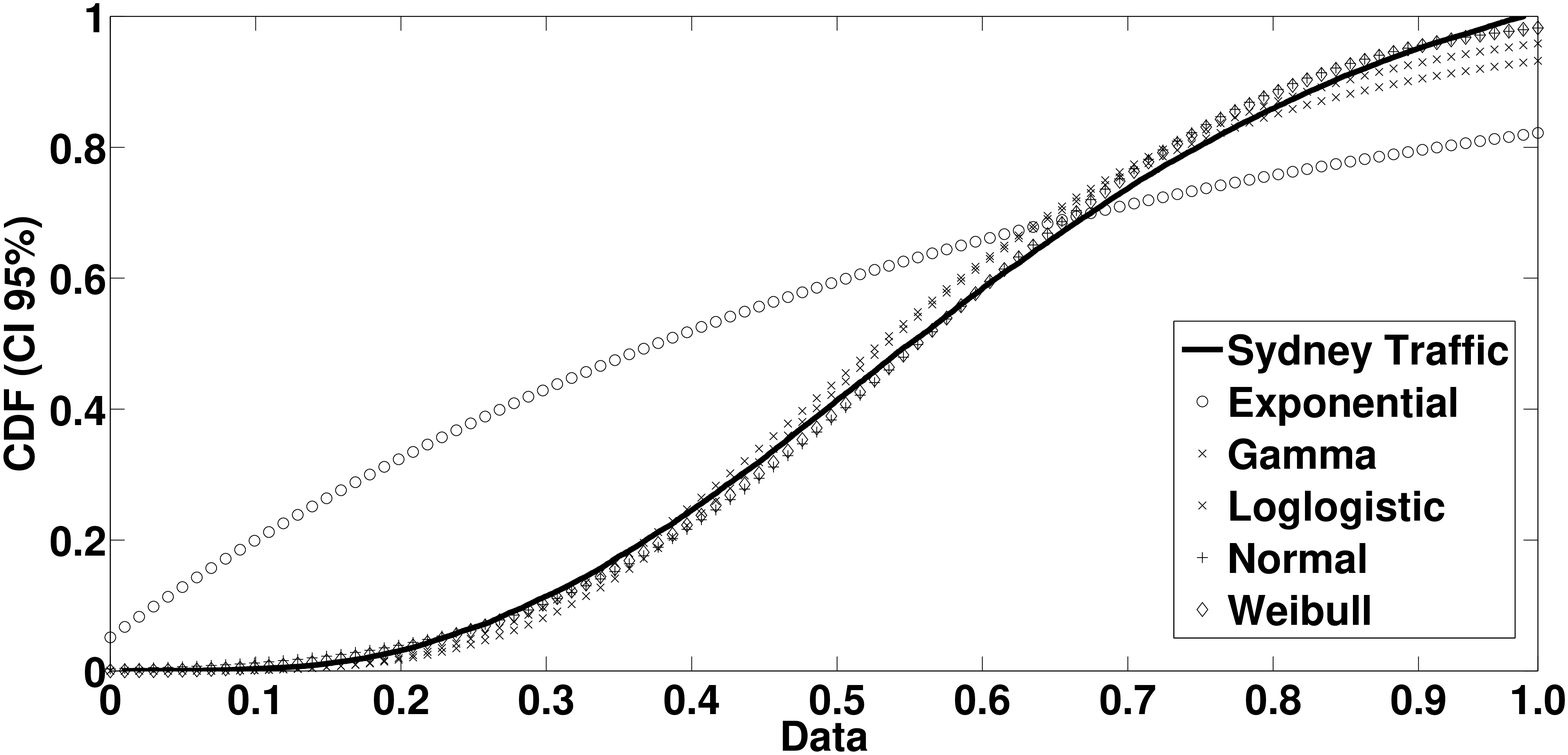}
   \label{fig:s2}
 }
 \subfigure[Sydney(H)]{
   \includegraphics [scale=.17]{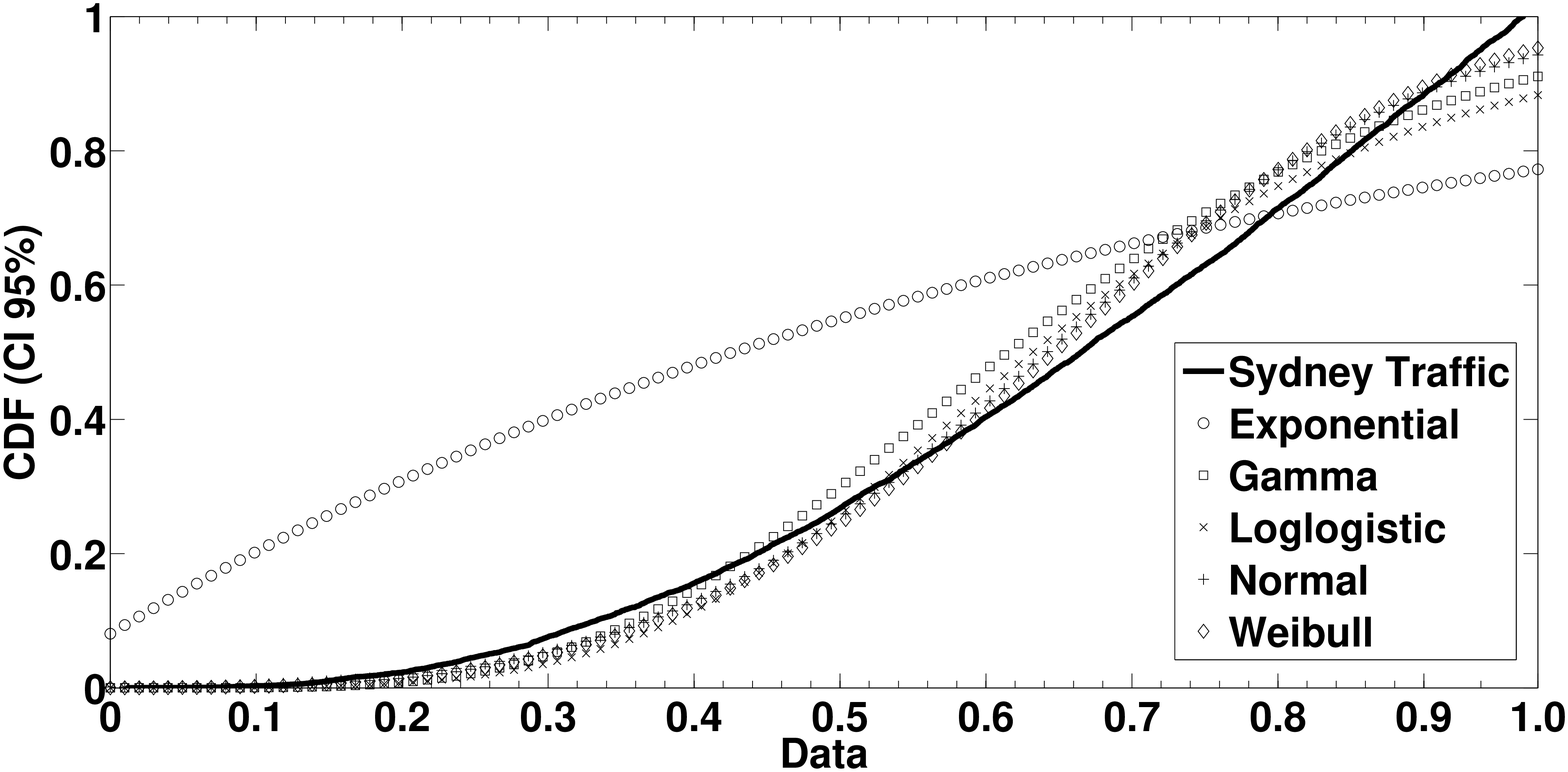}
   \label{fig:s3}
 }
 }
 \mbox{
 \subfigure[Toronto(L)]{
   \includegraphics [scale=.17]{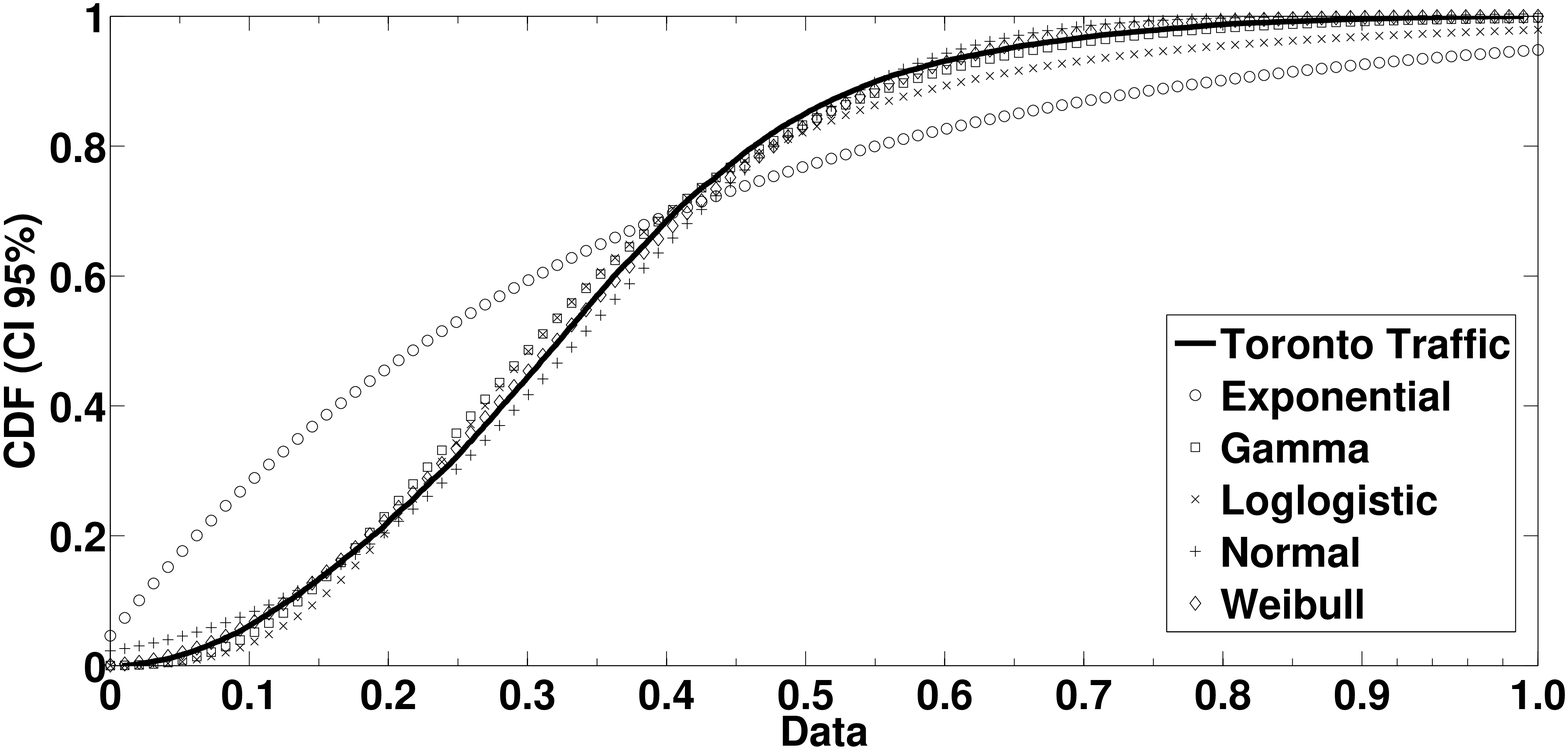}
   \label{fig:t1}
 }
 \subfigure[Toronto(M)]{
   \includegraphics [scale=.17]{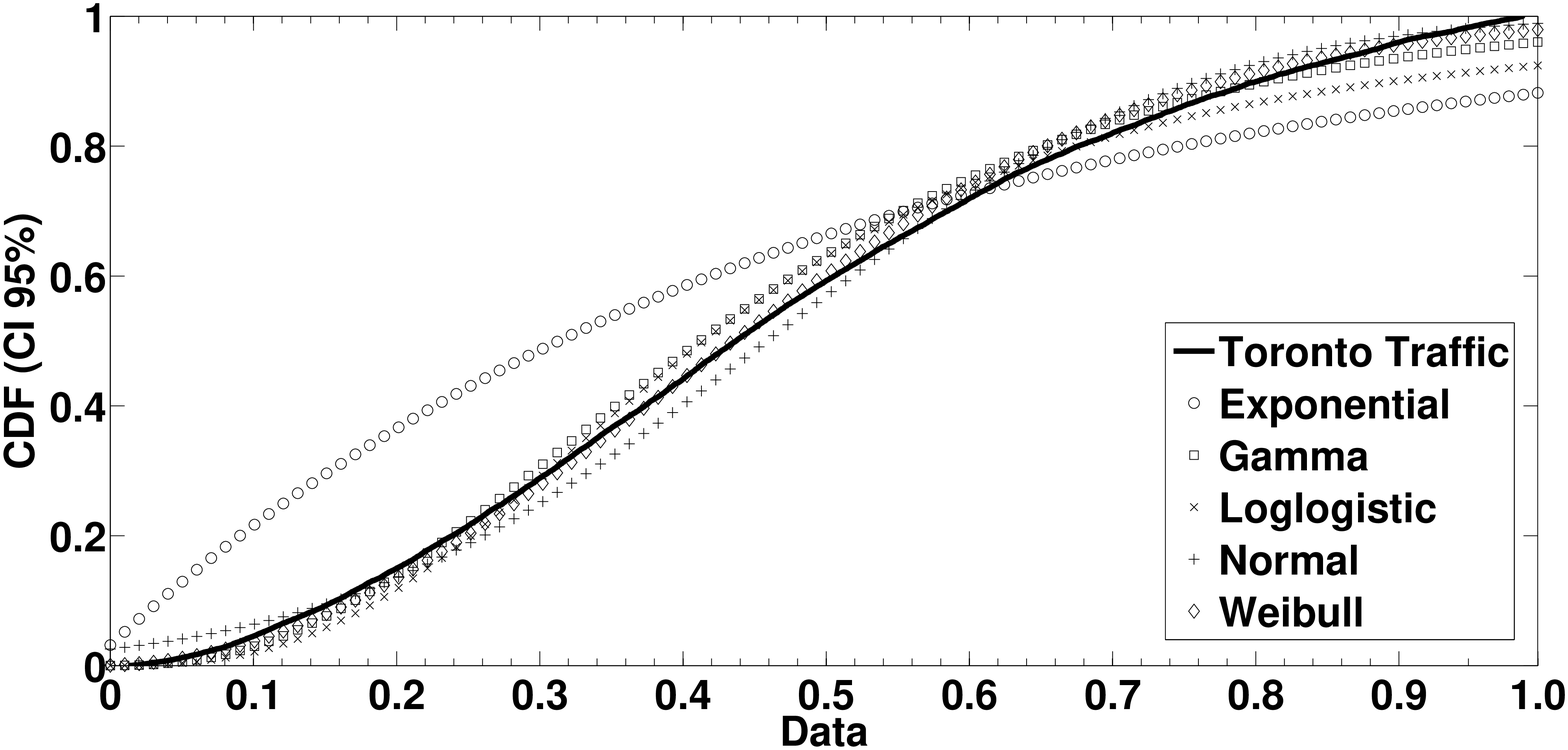}
   \label{fig:t2}
 }
 \subfigure[Toront(H)]{
   \includegraphics [scale=.17]{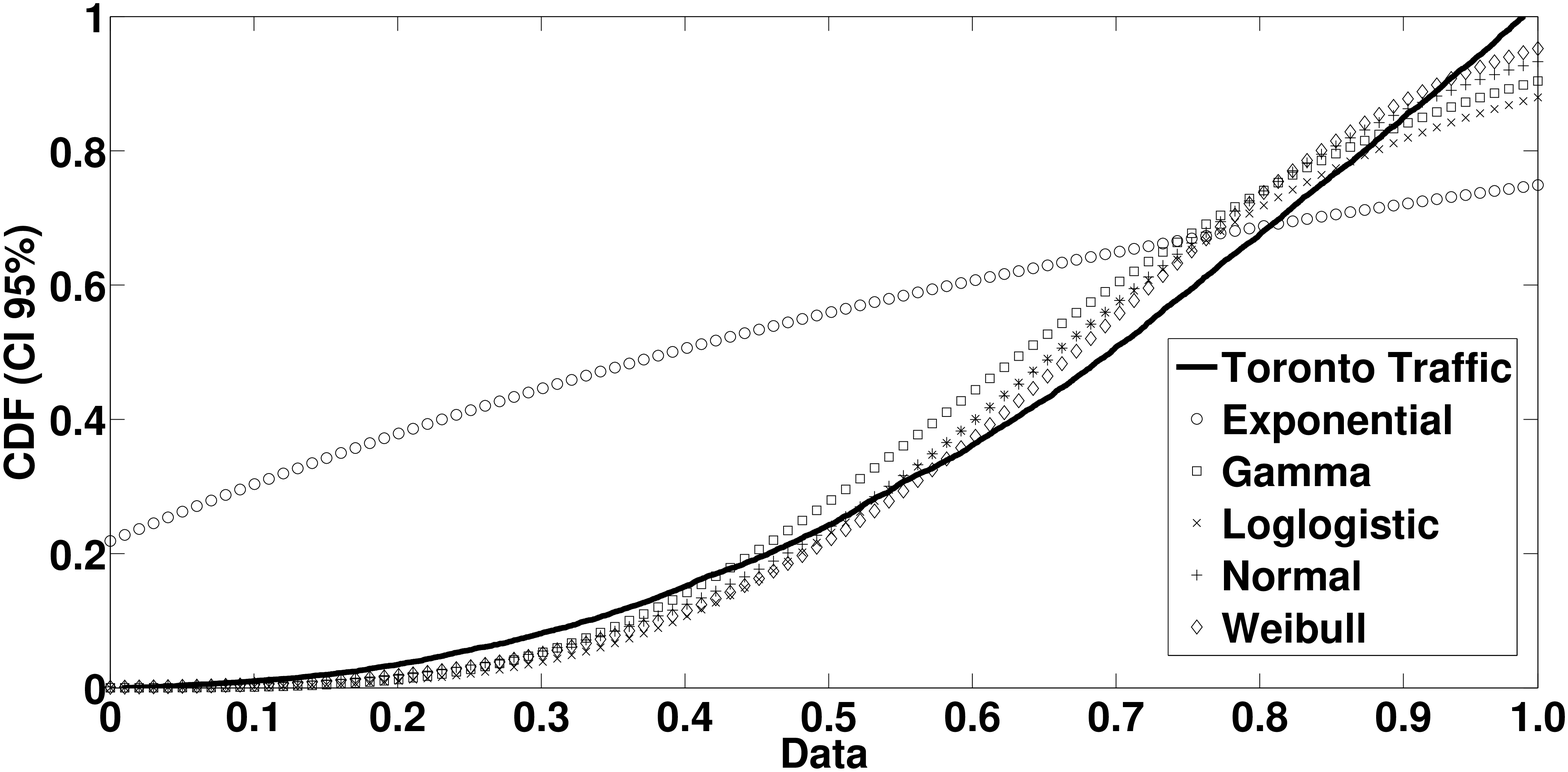}
   \label{fig:t3}
 }
}
\caption{Cumulative plot for three varying traffic intensities captured per city. The individual flows are characterized by the Low(L), Medium(M) and High(H) traffic intensities. }
\label{ccpdf}
\end{sidewaysfigure*}

\begin{table*}
\begin{center}
\scalebox{1}{
\begin{tabular}{|c|c|c|c|}\hline 
	\multicolumn{1}{|c|}{\textbf{City}} & \multicolumn{1}{|c|}{\textbf{1$^{st}$ Best Fit} }& \multicolumn{1}{|c|}{\textbf{2$^{nd}$ Best Fit}} & \multicolumn{1}{|c|}{\textbf{3$^{rd}$ Best Fit} }\\\hline 
	Connecticut & L[87\%] & G[11\%] & E[0.5\%] \\\hline 
	London & L[42\%] & G[39\%] & W[16\%] \\\hline 
	Sydney & L [62\%] & G[32\%] & N[2\%]  \\\hline 
	Toronto & G[46\%] & W[31\%] & L[21\%] \\\hline
	\multicolumn{4}{c}{E=Exponential. G=Gamma, L=Loglogistic, N= Normal, W=Weibull}
\end{tabular} }
\caption{Dominant distribution as Best Fits[By Ranking]}
\end{center}
\label{ddbf}
\end{table*}

\begin{table*}
\begin{center}
\scalebox{1}{
\begin{tabular}{|c|c|c|}\hline 
\textbf{City} & \textbf{$\leqslant$3\%} & \textbf{$\leqslant$5\% } \\\hline 
{Connecticut} & L[62\%], G[15\%], W[3\%]	  & L[94\%], G[44\%],  W[19\%]  \\ \hline 
{London} & G[34\%], L[34\%], W[10\%], N[0.5\%]&   L[82\%],  G[70\%], W[47\%], N[7\%] 	 \\\hline 
{Sydney} & L[88\%],  G[61\%], W[4\%],	N[2\%]&   L[98\%],    G[88\%],  W[44\%],  N[18\%] 	 \\\hline 
{Toronto} & G[75\%],  W[58\%], L[34\%] & G[94\%],  W[88\%], L[87\%], E[4\%], N[1\%]  \\	\hline 					
\end{tabular} }
\caption{Dominant distributions as Best Fits [By \% Deviation KS-Test.]}
\end{center}
\label{defaulttable}
\end{table*}


\section{Toward Realistic Vehicular Network Modeling}
As a first step toward realistic modeling of vehicular communication network, we focus on two studies of traffic arrival process in this paper: modeling the densities ($d$) against well known probability distributions and analyzing the typical traffic burstiness using self-similarity analysis. The objective of this study thus help to understand the underlying statistical patterns and  model the arrival processes. The models are selected based on their applicability in every day statistical analysis and by several iterations of modeling that showed the traffic closely follow (\textit{less deviation}) one or more of the discussed probability distributions. Due to page limit and as early study, in this section we will only present results from 4 represented cities (London, Sydney, Toronto, and Connecticut) with in total 458 cameras and 12 million images. An important and underlying fact about the traffic densities is the approximation to relative traffic on the roads. This assumption is different from counting cars using loop detectors or other sensors. As shown in the Fig.-4, we depict three traffic scenarios of varying intensities from low to fully congested intersection for the same camera as captured by the density parameter ($d$).

\begin{figure}[!h]
\centering
\epsfig{file=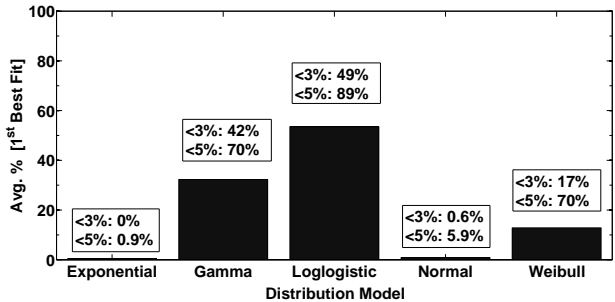, scale=0.2}
\caption{ The percentage of distribution that cover cameras from all four cities. The values in the box show percentage deviation error from empirical data. }
\label{wave}
\end{figure}

\subsection{Traffic Flow Characterization}
In order to investigate the nature of traffic we take a holistic approach to systematically extract individual and aggregate flows of the traffic densities from the images. Each individual flow constitutes a distribution of traffic densities that demonstrate the flow of traffic  as viewed from an individual camera. This helps us to better understand traffic intensity at a microscopic level of each intersection. The aggregate traffic combines the flows from all the camera in timely ordered fashion.  The main advantage from analyzing aggregate traffic is to understand the emergent properties and helps to model and profile the city and make intelligent guesses about different city based on this aggregate.  

On analyzing the traffic, an important activity to factorize the granularity of traffic for various purpose. For example,  hourly patterns provides a good estimate on the nature of congestions during morning and evening times which otherwise flow at individual density level may not depict. 
On the other hand, the finer granularity helps to understand sudden spikes in the traffic flow and congestion mitigation plan. In this work, we choose to look into all these patterns by modeling flows against well known probability distributions. Fig~\ref{imm} gives an example of the traffic density on hourly basis for one of the camera in Sydney. We can observe that there is in general high traffic density during the peak hours and low traffic density between 10am and 2pm (off peak time) which provides positive confirmation that our algorithm can effectively detect traffics.

Fig.~\ref{ccpdf} shows the cumulative density function of the traffic for three individual cameras in each city, with low, medium and high average traffic. We can see that traffic at individual cameras can vary a lot, but in general Log-Logistic, Gamma and Weibull distribution can capture some of the key features of the data. Log-logistic is the best approximation for the individual camera traffics in all the four cities, and we further shows the detail statistics of the fitting in Table-2 that best fits, which had shown least order of deviation against KS-test.

 In Table-3, we  measure the deviation from empirical data and sample the camera at 3\% and 5\% error levels. In Fig.-\ref{wave}, results show the average dominance of each of four distribution. We find that even on individual aggregation level, the loglogistic distribution provides a good estimate for empirical data. As evident, Loglogistic and Gamma closely matches the empirical data distribution. 
 
 Finally, Fig.~\ref{cpdf}  shows the cumulative statistics for the aggregated traffic for each city. We can observe that different cities have different aggregated traffic, for example we can see that London in general has more traffic than Connecticut.


\begin{sidewaysfigure*}
\centering
\epsfig{file=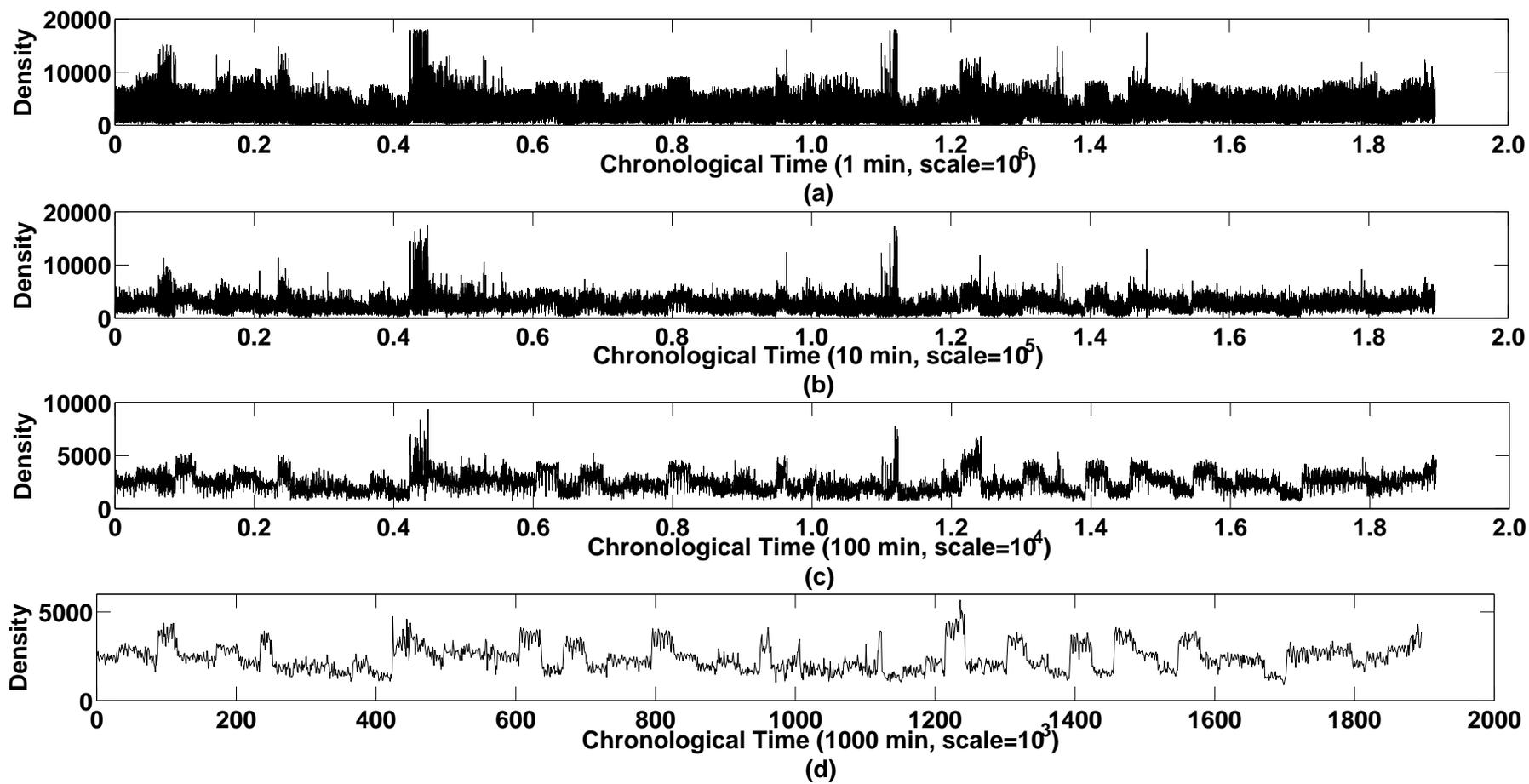,scale=0.6}
\caption{ Traffic density at different time scale on the Sydney dataset.}
\label{agg}
\end{sidewaysfigure*}


\subsection{Long Range Dependence}
In \cite{Leland:1994:SNE:178221.178222, citeulike:1044735}, authors demonstrate the existence of long range dependence and self-similar nature of ethernet traffic, which has serious implications on the design and analysis of computer networks.   Inspired by this study on the arrival process of ethernet packets in wired networks, we also characterize  the nature of vehicular traffic and investigate long range dependence.  Self-similarity means that aggregate traffic statistics show long range dependence and the correlation decays less than exponential. 
In Fig-\ref{agg}(a-d), we show  time series plots for four different chronological resolution of intervals for the city of Sydney. Initially, we plotted with  a time interval unit of one minute. The subsequent plots come from their previous plots but with one less order of resolution of time interval. A significant  burst is omni-present from finer to most abstract time resolutions. We also observed this behavior in other cities and we will further investigate in the future work by using different type of Hurst estimation\cite{1038171}.

\section{Conclusion and Future Work}
In this paper we introduced a novel method to collect large-scale vehicular network datasets using the always available online traffic webcams. These webcams are already deployed by governments, companies, or private and hence it is an inexpensive way for data collection. They provide 24 hours monitoring on the data collection points and have refresh rate as high as seconds, which is very desirable for fine grained data collection. We collected 7.5 TB of vehicular image data from more than 4,500 cameras distributed in 10 cites over 4 continents. We believe these large amount of data will be very important for mobile network researchers to understand the dynamics of the global cities and as a key step to realistic model vehicular communication networks. Our results strongly suggest a revisit to the general case of exponential pattern as modeling distribution for the vehicular traffic. Finally, the implication of long range dependence indicate the effect of traffic on the infrastructure of road networks. 

\section*{Acknowledgment}
We are thankful to Georgios Smaragdakis, Harold Chi Liu, Maria Gonzalez Garcia, Ranjan Pal and Shiva Sundaram for their insightful comments.

%
\bibliographystyle{abbrv}
\bibliography{sigproc}  

\begin{thebibliography}{10}

\bibitem{Bai:2009:SVV:1614269.1614278}
F.~Bai and B.~Krishnamachari.
\newblock Spatio-temporal variations of vehicle traffic in vanets: facts and
  implications.
\newblock In {\em Proceedings of the sixth ACM international workshop on
  VehiculAr InterNETworking}, VANET '09, pages 43--52, New York, NY, USA, 2009.
  ACM.

\bibitem{DBLP:conf/infocom/BaiSH03}
F.~Bai, N.~Sadagopan, and A.~Helmy.
\newblock Important: A framework to systematically analyze the impact of
  mobility on performance of routing protocols for adhoc networks.
\newblock In {\em INFOCOM}, 2003.

\bibitem{898398}
L.~Briesemeister, L.~Schafers, and G.~Hommel.
\newblock Disseminating messages among highly mobile hosts based on
  inter-vehicle communication.
\newblock In {\em Intelligent Vehicles Symposium, 2000. IV 2000. Proceedings of
  the IEEE}, pages 522 --527, 2000.

\bibitem{bychkovsky2006measurement}
V.~Bychkovsky, B.~Hull, A.~K. Miu, H.~Balakrishnan, and S.~Madden.
\newblock {A Measurement Study of Vehicular Internet Access Using In Situ Wi-Fi
  Networks}.
\newblock In {\em 12th ACM MOBICOM Conf.}, Los Angeles, CA, September 2006.

\bibitem{RobertE.Chandler03011958}
R.~E. Chandler, R.~Herman, and E.~W. Montroll.
\newblock Traffic dynamics: Studies in car following.
\newblock {\em OPERATIONS RESEARCH}, 6(2):165--184, 1958.

\bibitem{springerlink:elgammal}
A.~Elgammal, D.~Harwood, and L.~Davis.
\newblock Non-parametric model for background subtraction.
\newblock In D.~Vernon, editor, {\em Computer Vision — ECCV 2000}, volume
  1843 of {\em Lecture Notes in Computer Science}, pages 751--767. Springer
  Berlin / Heidelberg, 2000.

\bibitem{halati1997corsim}
A.~Halati, H.~Lieu, and S.~Walker.
\newblock {CORSIM-corridor traffic simulation model}.
\newblock In {\em Proceedings of the Traffic Congestion and Traffic Safety in
  the 21st Century Conference}, pages 570--576, 1997.

\bibitem{Hui:2010:PHM}
P.~Hui, R.~Mortier, M.~Pi\'{o}rkowski, T.~Henderson, and J.~Crowcroft.
\newblock Planet-scale human mobility measurement.
\newblock In {\em Proceedings of the 2nd ACM International Workshop on Hot
  Topics in Planet-scale Measurement}, HotPlanet '10, pages 1:1--1:5, New York,
  NY, USA, 2010. ACM.

\bibitem{Leland:1994:SNE:178221.178222}
W.~E. Leland, M.~S. Taqqu, W.~Willinger, and D.~V. Wilson.
\newblock On the self-similar nature of ethernet traffic (extended version).
\newblock {\em IEEE/ACM Trans. Netw.}, 2:1--15, February 1994.

\bibitem{1038171}
R.~Lienhart and J.~Maydt.
\newblock An extended set of haar-like features for rapid object detection.
\newblock In {\em Image Processing. 2002. Proceedings. 2002 International
  Conference on}, volume~1, pages I--900 -- I--903 vol.1, 2002.

\bibitem{Lownes:2006:VMS:1218112.1218368}
N.~E. Lownes and R.~B. Machemehl.
\newblock Vissim: a multi-parameter sensitivity analysis.
\newblock In {\em Proceedings of the 38th conference on Winter simulation}, WSC
  '06, pages 1406--1413. Winter Simulation Conference, 2006.

\bibitem{citeulike:1044735}
B.~Mandelbrot and J.~W. Van~Ness.
\newblock {Fractional Brownian Motions, Fractional Noises and Applications}.
\newblock {\em SIAM Review}, 10(4):422--437, 1968.

\bibitem{meng:864}
Q.~Meng and H.~L. Khoo.
\newblock Self-similar characteristics of vehicle arrival pattern on highways.
\newblock {\em Journal of Transportation Engineering}, 135(11):864--872, 2009.

\bibitem{1354509}
J.~Ott and D.~Kutscher.
\newblock Drive-thru internet: Ieee 802.11b for "automobile" users.
\newblock In {\em INFOCOM 2004. Twenty-third AnnualJoint Conference of the IEEE
  Computer and Communications Societies}, volume~1, pages 4 vol. (xxxv+2866),
  march 2004.

\bibitem{1400815}
M.~Piccardi.
\newblock Background subtraction techniques: a review.
\newblock In {\em Systems, Man and Cybernetics, 2004 IEEE International
  Conference on}, volume~4, pages 3099 -- 3104 vol.4, oct. 2004.

\bibitem{1040698}
J.~Singh, N.~Bambos, B.~Srinivasan, and D.~Clawin.
\newblock Wireless lan performance under varied stress conditions in vehicular
  traffic scenarios.
\newblock In {\em Vehicular Technology Conference, 2002. Proceedings. VTC
  2002-Fall. 2002 IEEE 56th}, volume~2, pages 743 -- 747 vol.2, 2002.

\bibitem{784637}
C.~Stauffer and W.~Grimson.
\newblock Adaptive background mixture models for real-time tracking.
\newblock In {\em Computer Vision and Pattern Recognition, 1999. IEEE Computer
  Society Conference on.}, volume~2, pages 2 vol. (xxiii+637+663), 1999.

\bibitem{10.1109/TPAMI.2006.104}
Z.~Sun, G.~Bebis, and R.~Miller.
\newblock On-road vehicle detection: A review.
\newblock {\em IEEE Transactions on Pattern Analysis and Machine Intelligence},
  28:694--711, 2006.

\bibitem{4346442}
N.~Wisitpongphan, F.~Bai, P.~Mudalige, V.~Sadekar, and O.~Tonguz.
\newblock Routing in sparse vehicular ad hoc wireless networks.
\newblock {\em Selected Areas in Communications, IEEE Journal on}, 25(8):1538
  --1556, oct. 2007.

\bibitem{Yeo:2006:CCR:1129582.1129588}
J.~Yeo, D.~Kotz, and T.~Henderson.
\newblock Crawdad: a community resource for archiving wireless data at
  dartmouth.
\newblock {\em SIGCOMM Comput. Commun. Rev.}, 36:21--22, April 2006.

\bibitem{Zhang:2007:SBD:1287853.1287876}
X.~Zhang, J.~Kurose, B.~N. Levine, D.~Towsley, and H.~Zhang.
\newblock Study of a bus-based disruption-tolerant network: mobility modeling
  and impact on routing.
\newblock In {\em Proceedings of the 13th annual ACM international conference
  on Mobile computing and networking}, MobiCom '07, pages 195--206, New York,
  NY, USA, 2007. ACM.

\end{thebibliography}
%
%

\end{document}